\documentclass[12pt]{iopart}
\usepackage{iopams}
\usepackage{epsfig}
\usepackage{graphicx, subfigure}
\usepackage{multirow}
\usepackage{hyperref}
\usepackage{cite}
\usepackage{color}

\begin{document}

\title[quantum-classical approach]{Quantum-quasiclassical analysis of center-of-mass nonseparability in hydrogen atom stimulated by strong laser fields}

\author{Vladimir S. Melezhik$^{1,2}$}
\address{$^1$ Bogoliubov Laboratory of Theoretical Physics, Joint Institute for Nuclear Research, Dubna, Moscow Region 141980, Russian Federation}
\address{$^2$ Dubna State University, 19 Universitetskaya St., Moscow Region 141982, Russian Federation}

\eads{\mailto{melezhik@theor.jinr.ru}}

\vspace{10pt}



\begin{abstract}
We have developed a quantum-quasiclassical computational scheme for quantitative treating of the nonseparable quantum-classical dynamics of the 6D hydrogen atom in a strong laser pulse. In this approach, the electron is treated quantum mechanically and the center-of-mass (CM) motion classically. Thus, the Schr\"odinger equation for the electron and the classical Hamilton equations for the CM variables, nonseparable due to relativistic effects stimulated by strong laser fields, are integrated simultaneously.  In this approach, it is natural to investigate the idea of using the CM-velocity spectroscopy as a classical ``build-up'' set up for detecting the internal electron quantum dynamics. We have performed such an analysis using the hydrogen atom in linearly polarized laser fields as an example and found a strong correlation between the CM kinetic energy distribution after a laser pulse and the spectral density of electron kinetic energy. This shows that it is possible to detect the quantum dynamics of an electron by measuring the distribution of the CM kinetic energy.
\end{abstract}



\section{Introduction}
In the works~\cite{Melezhik2000,Melezhik2004,Melezhik2019,Melezhik2021} the efficient quantum-quasiclassical computational scheme was developed that was successfully applied to calculate various few-body processes and has made it possible to resolve a number of topical problems in atomic~(see \cite{Melezhik2021} and references therein), mesoatomic, and nuclear physics. In this approach, the few-body quantum problem is reduced to the simultaneous integration of a system of coupled quantum and classical equations: the time-dependent Schr\"odinger equation that describes the quantum dynamics of  slow light particles and the classical Hamilton equations that describe the remaining variables of heavy fast particles. The key idea of this semiclassical approach goes back to works~\cite{Method1,Method2,Method3} where it was applied to the molecular dynamics. Recently~\cite{Melezhik2019}, the method has been extended and adapted to the quantitative description of pair collisions of light slow Li atoms with heavy Yb$^+$ ions in the confined geometry of the hybrid atom-ion trap. On the basis of these calculations,  a new method for sympathetic cooling of ions in the RF Paul trap was proposed~\cite{Melezhik2021}: to use cold buffer atoms for this purpose in the region of atom-ion confinement-induced resonance. This approach also made it possible to perform calculations of the breakup cross sections in the low-energy region (up to 10 MeV/nucleon), inaccessible so far to other methods, for the $^{11}$Be breakup on a heavy target~\cite{Valiolda2022}. Here, we extend the method for quantitative analysis of the nonseparable quantum-classical dynamics of the 6D hydrogen atom in a strong laser pulse.

In a recent paper by Patchkovskii et. al.~\cite{Patchkovskii}, it was found that the non-dipole coupling in the hydrogen atom between the center-of-mass (CM) and the electron motions induced by external strong laser fields leads to a correlation between the CM-velocity distribution after the pulse action with the population of electron states induced by the field. Therefore, the authors of this paper suggest using this effect for detection of the internal electron quantum dynamics with CM-velocity spectroscopy. Moreover, such a ``device'' has an advantage that it can be considered as a classical spectroscope for investigation of the quantum dynamics inside the atom.

The problem of accurate treatment of the non-dipole coupling of the CM- and relative electron-proton motions in a hydrogen atom effected by a strong laser field is rather challenging due to the 6D dimensionality of the problem and the smallness of the coupling effect which is of the order $\sim 1/c \sim 1/137$. In the work~\cite{Patchkovskii}, the original 6D problem was replaced by the effective 3D Schr\"odinger equation by placing the atom in an artificial trap (spherical harmonic oscillator) with an energy corresponding to the energy of its thermal motion. After averaging the motion of the atom over the states of the harmonic oscillator, they reduced the original 6D problem to the effective 3D Schr\"odinger equation for the electron. In the present work, we propose to avoid this drawback by applying our quantum-quasiclassical~\cite{Melezhik2000,Melezhik2004,Melezhik2019,Melezhik2021} approach to the 6D problem of a hydrogen atom in a laser field in which the electron motion is treated quantum mechanically  and the CM variables classically. In this approach, simultaneously with the time-dependent Schr\"odinger equation for the electron wave function we integrate a set of classical Hamilton equations describing CM-variables.

With this approach, we have calculated the time dynamics of the electron and CM- variables as well as the spectral densities of the electron and CM kinetic energies after the linearly polarized laser pulses of rather high intensity ($10^{14}\frac{W}{cm^2}$) and broad region ($90-800 nm$)of the wave lengths. The populations of the low-lying hydrogen energy levels stimulated by the laser fields have also been evaluated. These calculations show a strong correlation between the atom kinetic energy distribution after the pulse and the electron spectral density inside the atom. This fact supports the idea~\cite{Patchkovskii} of using the CM- velocity spectroscopy as a classical spectroscope for detecting the quantum electron dynamics inside the atom.

In the next section, we derive equations of the quantum-quasiclassical approach as applied to the hydrogen atom in strong linearly polarized laser fields. The results of the calculations with this method of the hydrogen atom dynamics in strong laser fields and the discussion are given in the third section. The last section is devoted to short conclusion.

\section{Treating 6D hydrogen atom in strong laser field with quantum-quasiclassical approach}
We investigate the dynamics of the 6D hydrogen atom in a strong laser field linearly polarized along x-axis (atomic units $e^2=m_e=\hbar=1$ are used hereinafter except where otherwise noted) with the electric ${\bi E}$ and magnetic ${\bi B}$ fields
\begin{eqnarray}\label{field}
{\bi E}(\omega t) = E(\omega t){\bi n}_x \,\,,\,\,\,\,\, {\bi B}(\omega t) = \frac{1}{c} E(\omega t){\bi n}_y \,\,,
\end{eqnarray}
where
\begin{eqnarray}\label{field0}
E(\omega t) = E_0 f(t)\cos(\omega t- kz)
\end{eqnarray}
and
\begin{eqnarray}\label{field1}
f(t) = \cos^2(\frac{\pi t}{n_TT}) \,\,, \,\,\,\, -n_TT/2\leq t \leq n_TT/2\,\,.
\end{eqnarray}
The pulse duration determined in this way includes $n_T$ optical cycles with cycle period $T=2\pi/\omega$.
Here, ${\bi n}_x$ and ${\bi n}_y$ are unit vectors along the x and y axes, $1/c=\alpha=1/137$ is the fine structure constant, $k=\omega/c$ is the wave number and the strength of the laser field $E_0$ defined by the field intensity $I = \epsilon_0 c E_0^2/2=I_0 E_0^2$, where $\epsilon_0$ is the vacuum permittivity and $I_0=3.51 \times 10^{16}\frac{W}{cm^2}$. All calculations were performed for intensity $I=10^{14}\frac{W}{cm^2}$. The laser field frequency $\omega$ is connected by known relation $\omega=2\pi c/\lambda$ with the wave length, which in our calculation was alternating in the region $90nm \leq \lambda \leq 800nm$. Pulse propagates along the z-axis.

A vast majority of calculations of the dynamics of the hydrogen atom in laser fields were performed in the dipole approximation for the atom-field interaction potential $V({\bi r},t)= E_0 f(t)\cos(\omega t)x$ (where $x=x_e-x_p$ and $x_e$ and $x_p$ are the variables of electron and proton, respectively), in which the magnetic component ($\sim\alpha$) in equations (\ref{field}) and the spatial dependence in the propagation direction of the pulse ($\sim kz =\alpha\omega z$) in (\ref{field0}) are neglected. Accounting for the magnetic component in (\ref{field}) and the spatial dependence of the laser field in (\ref{field0}), i.e. going beyond the dipole approximation, leads to the following modification of the interaction potential (see Appendix A)
\begin{eqnarray}\label{interaction}
V({\bi r},{\bi R},t) = V_1({\bi r},t) + V_2({\bi r},{\bi R},t)\,\,,
\end{eqnarray}
where
\begin{eqnarray}\label{interaction1}
V_1({\bi r},t) = E_0f(t)\{\cos(\omega t)x +\alpha[\cos(\omega t)\hat{l}_y + \omega\sin(\omega t)xz]\}\,\,,
\end{eqnarray}
and
\begin{eqnarray}\label{interaction2}
V_2({\bi r},{\bi R},t) = \alpha E_0f(t)\{\cos(\omega t)[Z\hat{p}_x-X \hat{p}_z] + \omega\sin(\omega t)[xZ + zX]\}\,\,.
\end{eqnarray}
This potential is written in the CM ${\bi R}=(X,Y,Z)$ and relative ${\bi r}= (x,y,z)$ variables, where $\hat{l}_y = z\hat{p}_x-x\hat{p}_z$ is the y-th component of the operator of the electron angular momentum with respect to the proton. In deriving these formulas, we neglected the terms $\sim \alpha^2$ and $\sim 1/M = 1/(m_p+m_e)$ and higher orders (Appendix A).
Then, the total Hamiltonian of the system takes the form
\begin{eqnarray}\label{hamiltonian}
H({\bi r},{\bi R}, t) = \frac{{\bi P}^2}{2M} + h_0({\bi r}) +V_1({\bi r},t) + V_2({\bi r},{\bi R},t)\,\,,
\end{eqnarray}
where
\begin{eqnarray}\label{hamiltonian0}
h_0({\bi r}) = \frac{\hat{\bi p}^2}{2\mu} -\frac{1}{r} \,\,,
\end{eqnarray}
$\hat{\bi p}$ is the momentum operator of the relative motion of the electron with respect to the proton, ${\bi P}$ is the momentum of the CM, $\mu = m_e m_p/(m_e+m_p)$ is the reduced mass of the atom and $M=m_e+m_p$.

The importance of non-dipole effects was recognized long ago~\cite{Reiss1,Kylstra,Hemmers,Forre0}, and by now the influence of these effects on various atomic processes in strong laser fields has been studied quite intensively (see, for example,~\cite{Reiss2,Smeenk,Ludwig,Chelkowski,Forre1,Kleiber,Ilchen,Maurer,Forre2} and references therein). However, non-separability of the CM motion has not received much attention so far~\cite{Patchkovskii}, as we suppose, due to the computational complexity of the problem. Here, we eliminate this gap to some extent by proposing a computational method for the quantitative analysis of the nonseparable quantum-classical dynamics of the 6D hydrogen atom in a strong laser pulse, taking into account the motion of the CM and its coupling (\ref{interaction2}) with the electron motion.

Since in our problem ${\bi P}=M{\bi V}\gg \mu{\bi v}$, we can consider the motion of a heavy atom as a whole classically (CM- motion) and the motion of a light electron relative to a proton in an atom is quantum. It also justified by the well-known fact that the classical model of Maxwell-Boltzmann ideal gas perfectly describes gas laws down to sufficiently low temperatures. This allows us to apply here the algorithm of the quantum-quasiclassical approach~\cite{Melezhik2000,Melezhik2004,Melezhik2019,Melezhik2021} and, following this scheme, reduce the original problem of the hydrogen atom in a laser field to the integration of the following system of coupled equations
\begin{eqnarray}\label{Schrodinger}
i\hbar\frac{\partial}{\partial t}\psi({\bi r},t) = \{h_0({\bi r})+V_1({\bi r},t) + V_2({\bi r},{\bi R}(t),t)\}\psi({\bi r},t)\,\,,
\end{eqnarray}
\begin{eqnarray}\label{newton1}
\frac{d}{dt}{\bi P}(t) = -\frac{\partial}{\partial {\bi R}}H_{cl}({\bi R}(t),{\bi P}(t))\,\,,
\end{eqnarray}
\begin{eqnarray}\label{newton2}
\frac{d}{dt}{\bi R}(t) = -\frac{\partial}{\partial {\bi P}}H_{cl}({\bi R}(t),{\bi P}(t))\,\,,
\end{eqnarray}
with the effective Hamiltonian
\begin{eqnarray}\label{newton3}
H_{cl}({\bi R},{\bi P}) = \frac{{\bi P}^2}{2M} + \langle\psi({\bi r},t)|V_2({\bi r},{\bi R}(t),t)|\psi({\bi r},t)\rangle\,\,.
\end{eqnarray}
To solve problem (\ref{Schrodinger}-\ref{newton3}), it is necessary to set the initial conditions for $t=-n_TT/2$
\begin{eqnarray}\label{initial}
\psi({\bi r},t=-n_TT)= \phi_{nlm}({\bi r})\,\,,
\end{eqnarray}
\begin{eqnarray}\label{initial1}
{\bi R}(t=-n_TT)={\bi R}_0\,\,,\,\,{\bi P}(t=-n_TT)={\bi P}_0\,\,,
\end{eqnarray}
and integrate simultaneously the system of the coupled equations~(\ref{Schrodinger}-\ref{newton3}), where $\phi_{nlm}({\bi r})$ is the hydrogen atom wave function of the bound state $|nlm\rangle$.

A feature of problem (9-14) is the presence in it of the characteristic frequencies of the laser pulse and transitions between the levels of atoms, which are very different from each other. Therefore, the algorithm used for the numerical integration of the system (9-14) must be stable in a wide frequency range. To integrate the time-dependent 3D Schr\"odinger equation~(\ref{Schrodinger}) we apply our recently developed algorithm~\cite{Shadmehri} based on splitting-up method with a 2D discrete-variable representation (DVR)~\cite{Melezhik97,Melezhik99}. It was successfully used to calculate, in the dipole approximation, the excitation and ionization of a hydrogen atom by a strong elliptically polarized laser field ($10^{14}\frac{W}{cm^2}$). Simultaneously to the forward in time propagation $t_n \rightarrow t_{n+1} = t_n + \Delta t$ of the electron wave-packet $\psi({\bi r}, t_n ) \rightarrow
\psi({\bi r}, t_{n+1})$ when integrating the time-dependent Schrödinger
equation (\ref{Schrodinger}), we integrate the Hamilton equations of motion
(\ref{newton1},\ref{newton2}) with the  St\"ormer-Verlet method~\cite{Wanner} adapted in~\cite{Melezhik2019,Melezhik2021} for quantum-quasiclassical case:
\begin{eqnarray}
{\bi P}(t_n+\frac{\Delta t}{2})= {\bi P}(t_n) -\frac{\Delta t}{2}\frac{\partial}{\partial {\bi R}} H_{cl}({\bi P}(t_n+\frac{\Delta_t}{2}), {\bi R}(t_n))\,\,,\nonumber
\end{eqnarray}
\begin{eqnarray}
{\bi R}(t_n+\Delta t) = {\bi R}(t_n) + \frac{\Delta t}{2}\left\{\frac{\partial}{\partial {\bi P}} H_{cl}({\bi P}(t_n+\frac{\Delta t}{2}),{\bi R}(t_n))\right.\nonumber \\
\left.+\frac{\partial}{\partial {\bi P}}H_{cl}({\bi P}(t_n+\frac{\Delta t}{2}),{\bi R}(t_n+\Delta t))\right\}\,,\nonumber
\end{eqnarray}
\begin{equation}\label{SW}
{\bi P}(t_n+\Delta t) = {\bi P}(t_n+\frac{\Delta t}{2}) - \frac{\Delta t}{2}\frac{\partial}{\partial {\bi R}}H_{cl}({\bi P}(t_n+\frac{\Delta t}{2}),{\bi R}(t_n+\Delta t))\,\,.
\end{equation}

We calculate $\psi({\bf r},t)$ and ${\bi R}(t)$,${\bi P}(t)$ and average values of the kinetic energy of the atom as a whole $\langle|E_{kin}|\rangle$ and the electron kinetic energy $\langle|E_{kin}^{(el)}|\rangle$ after the pulse end
\begin{eqnarray}\label{ekinM}
\langle|E_{kin}|\rangle = \frac{1}{T_{out}-T_{in}}\int_{T_{in}}^{T_{out}}\frac{{\bi P}^2(t)}{2M}dt\sim\int_{-\infty}^{\infty}[\sum_{s=x,y,z}|P_k(\omega)|^2] d\omega\,,
\end{eqnarray}
\begin{eqnarray}\label{ekinE}
\langle|E_{kin}^{(el)}|\rangle = \frac{1}{T_{out}-T_{in}}\int_{T_{in}}^{T_{out}}\frac{{\bi p}^2(t)}{2\mu}dt\sim\int_{-\infty}^{\infty}[\sum_{s=x,y,z}|p_k(\omega)|^2]d\omega\,,
\end{eqnarray}
as well as the spectral densities $|P_s(\omega)|^2$ and $|p_s(\omega)|^2$ of the atom and electron kinetic energies, respectively (where $s=x,y,z$, $T_{in}=-n_TT$ and $T_{out}=(n_T+1)T$). Here
\begin{eqnarray}\label{ekino}
P_s(\omega) = \int_{T_{in}}^{T_{out}} P_s(t) e^{i\omega t}dt\,\,
\end{eqnarray}
and
\begin{eqnarray}\label{ekinoe}
p_s(\omega) = \int_{T_{in}}^{T_{out}} \langle|p_s(t)|\rangle e^{i\omega t}dt\,\,
\end{eqnarray}
where the distributions of $p_s(\omega)$ are expressed in terms of the instantaneous values $\langle|\hat{p}_s(t)|\rangle$ of the electron momentum, calculated simultaneously with integrating system (\ref{Schrodinger}-\ref{newton2})
\begin{eqnarray}\label{ekint}
\langle|p_s(t)|\rangle = \int \psi^*({\bi r},t) \hat{p}_s \psi({\bi r},t)d{\bi r}\,\,.
\end{eqnarray}

\section{Electron and CM- dynamics of hydrogen atom in strong laser fields}

Using the computational scheme presented in the previous section, we have calculated the dynamics of the hydrogen atom in a linearly polarized laser field of $10^{14}\frac{W}{cm^2}$ intensity for three wavelengths $800nm$, $400nm$ and $90nm$. The transition from longer to shorter waves makes it possible to analyse the effect of amplifying the coupling between the internal motion of the electron and the CM- motion, since in this case the radiation frequency $\omega=2\pi c/\lambda$ increases and, as a consequence, the cross term $V_2({\bi r}, {\bi R}(t),t)$ (\ref{interaction2}) in the Hamiltonian (\ref{hamiltonian}) of the problem is amplified.
All calculations were performed for the hydrogen atom in its initial state ($t=-n_TT$) in the ground state $\phi_{100}({\bi r})$ at the origin of the coordinates (${\bi R}_0=0$) with zero momentum (${\bi P}_0=0$).

Figure 1 shows the results of the calculation of ${\bi R}(t)$, ${\bi P}(t)$ and $\langle|\hat{{\bi p}}(t)|\rangle$ for $\lambda=800 nm$ ($\omega=0.057$). The calculations were carried out for $\Delta t=T/4400=\pi/(2200\omega)$ on a radial grid $0\leq r \leq 500$ with $N_r=2000$ grid points and an angular grid with $N_{\theta}=11$ and $N_{\phi}=11$ Guassian nodes in $\theta$ and $\phi$ variables, respectively, which give a convergent result. The construction of a spatial grid and the method for analyzing convergence in integrating Eq.(9) on a sequence of refining grids are described in~\cite{Shadmehri}.
The calculated curves $X(t)$, $Z(t)$, $P_x(t)$ and $P_z(t)$ show that the atom is accelerated under the action of the laser field in accordance with the known experimental results~\cite{Eichmann}. Moreover, the acceleration in the direction of pulse propagation ($z$-axis) is three orders of magnitude higher than the acceleration in the direction of polarization ($x$-axis). Interaction with the laser field also leads to electron acceleration (see curves $\langle|p_x(t)|\rangle$ and $\langle|p_z(t)|\rangle$). Moreover, its amplitude of oscillations in the direction of propagation of the laser pulse is two orders of magnitude smaller than the amplitude of oscillations in the direction of laser polarization and significantly lags in time. It should also be noted that at the initial stage of the interaction of an atom with a laser field ($t < 0$), electron oscillations are determined by the laser frequency $\omega=0.057$, but starting from times approaching to $t\sim 0$, high-frequency components appear in the electron momentum $\langle|p_x(t)|\rangle$, which we associate with the excitation of low-lying levels of the atom, i.e. the frequencies $\Omega \sim \frac{1}{2}-\frac{1}{8} \sim 0.375 \gg \omega = 0.057$ related with the transition $n=1\rightarrow n'=2,3 ..$~. This high-frequency effect is also visible in $\langle|p_z(t)|\rangle$ and is only slightly noticeable in the $P_x(t)$ and $P_z(t)$ component of the atomic momentum.
\begin{figure}[!]
\begin{center}
\begin{subfigure}{
\includegraphics[width=0.45\columnwidth]{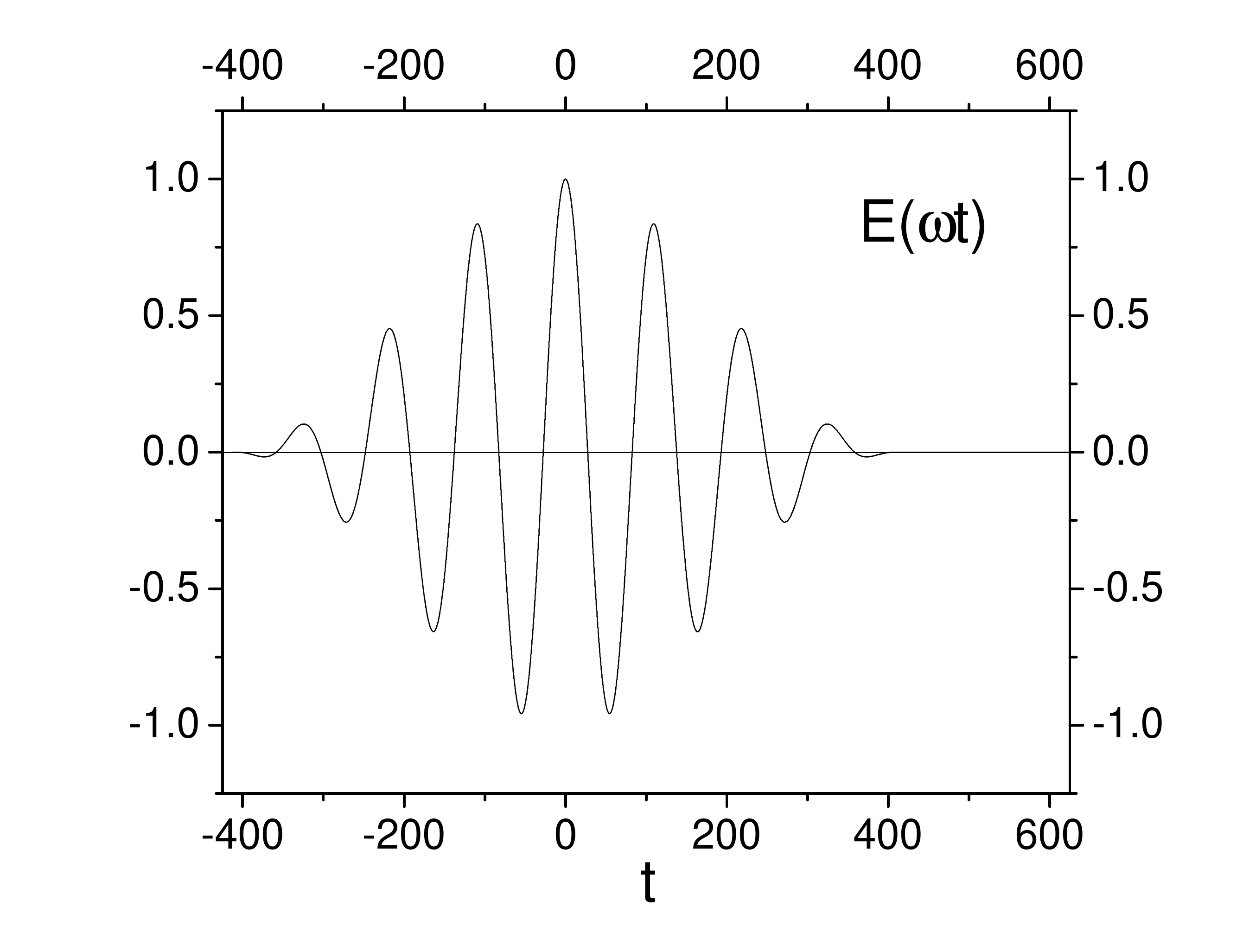} \includegraphics[width=0.45\columnwidth]{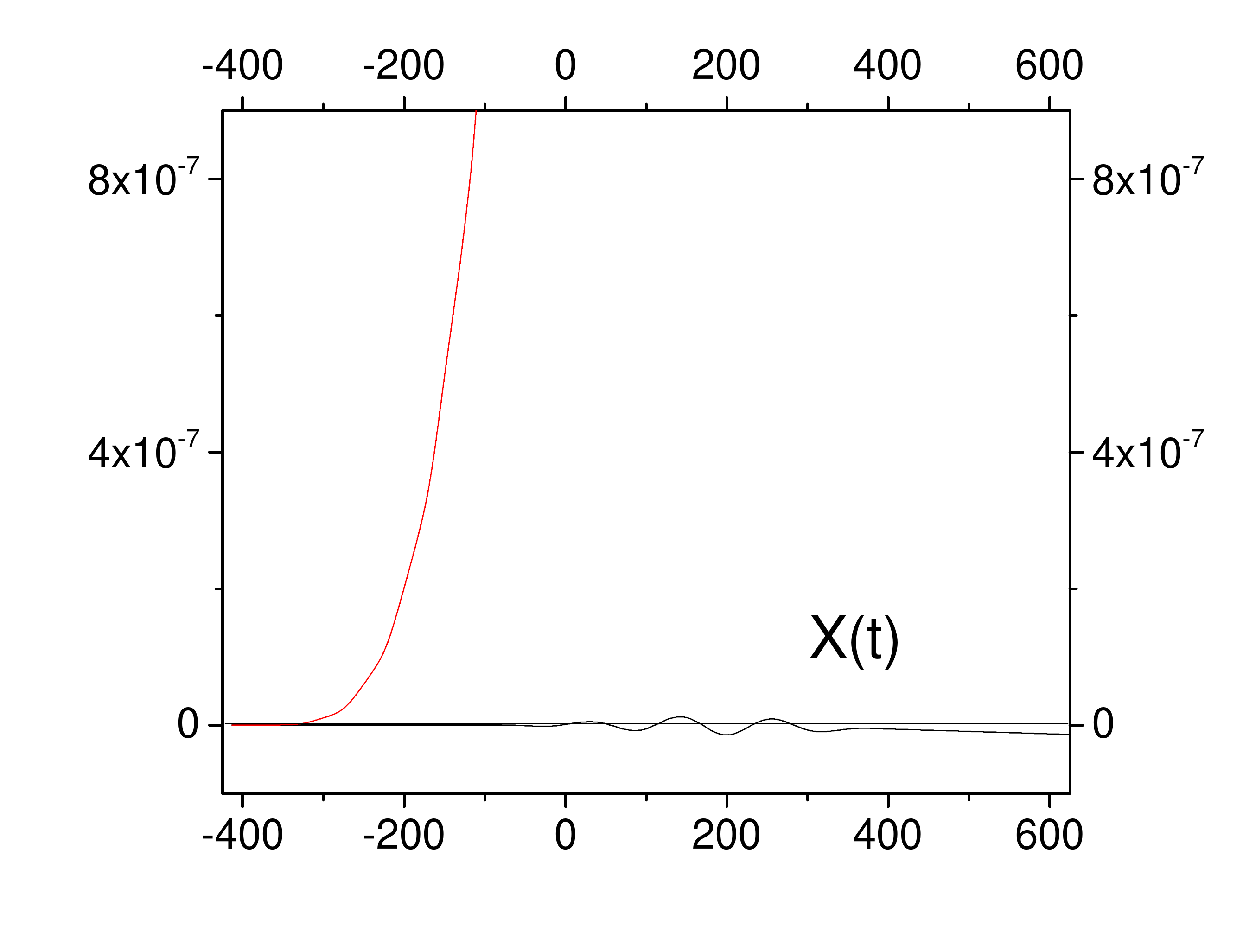}}
\end{subfigure}\\
\begin{subfigure}{
\includegraphics[width=0.45\columnwidth]{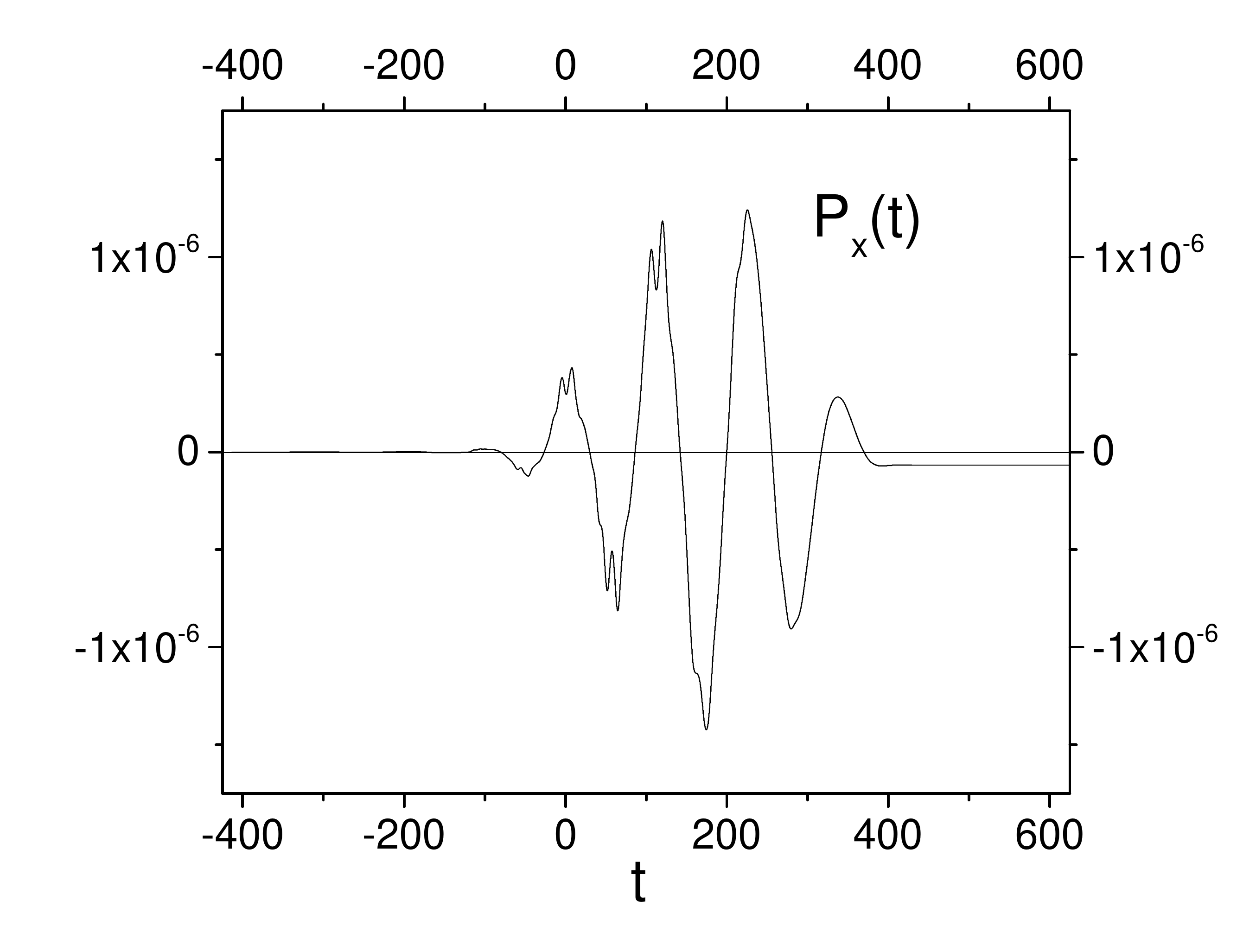} \includegraphics[width=0.45\columnwidth]{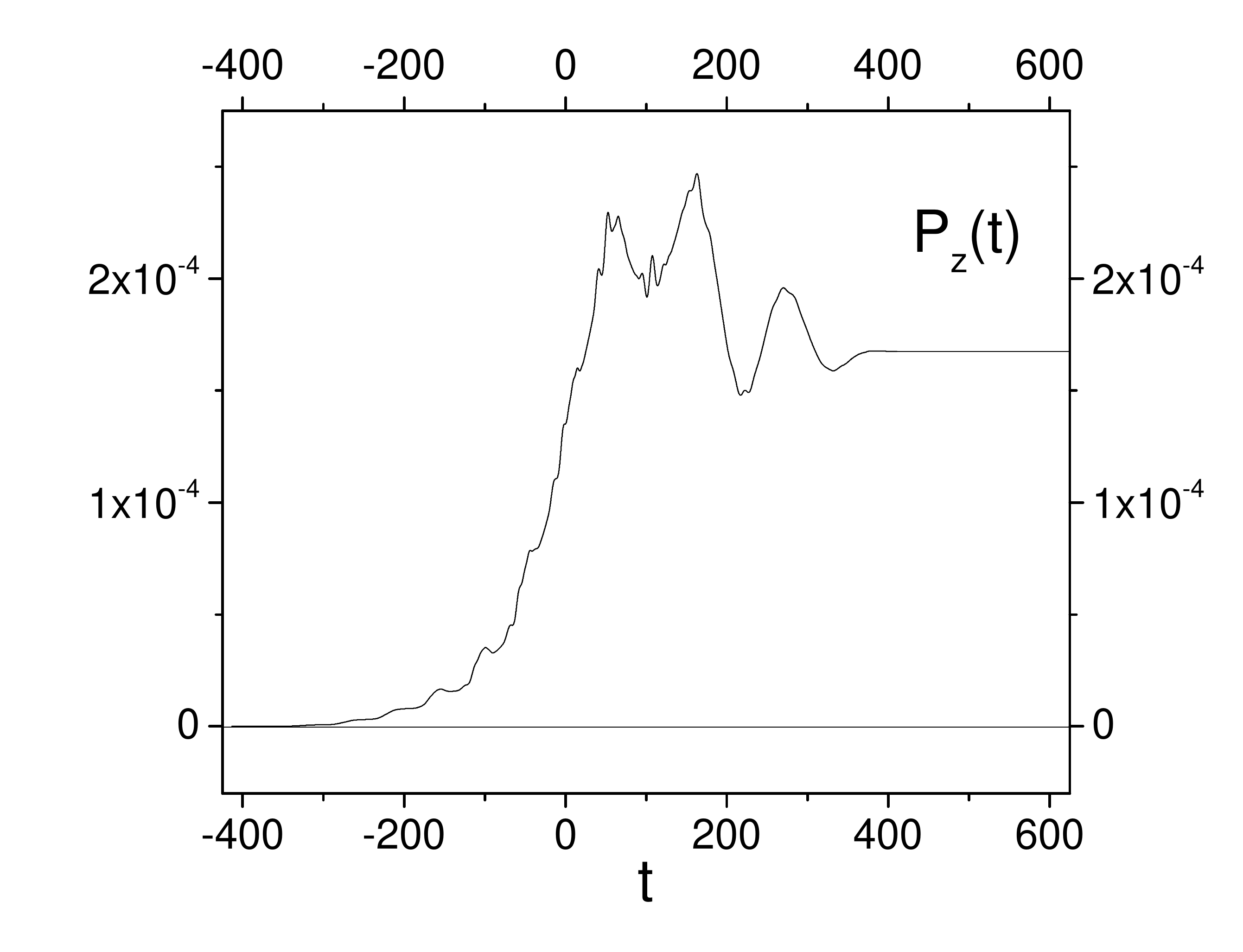}}
\end{subfigure}
\begin{subfigure}{
\includegraphics[width=0.45\columnwidth]{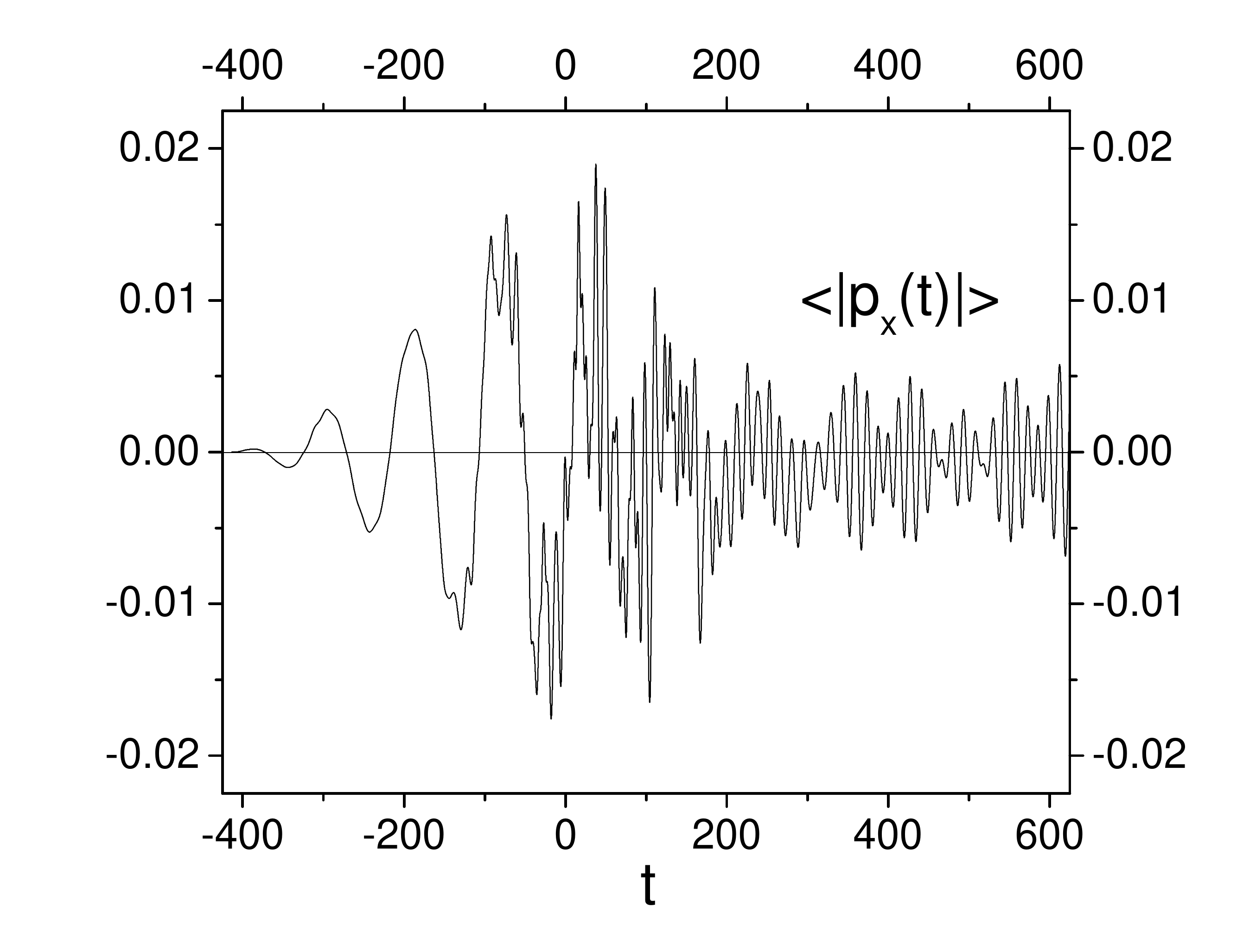} \includegraphics[width=0.45\columnwidth]{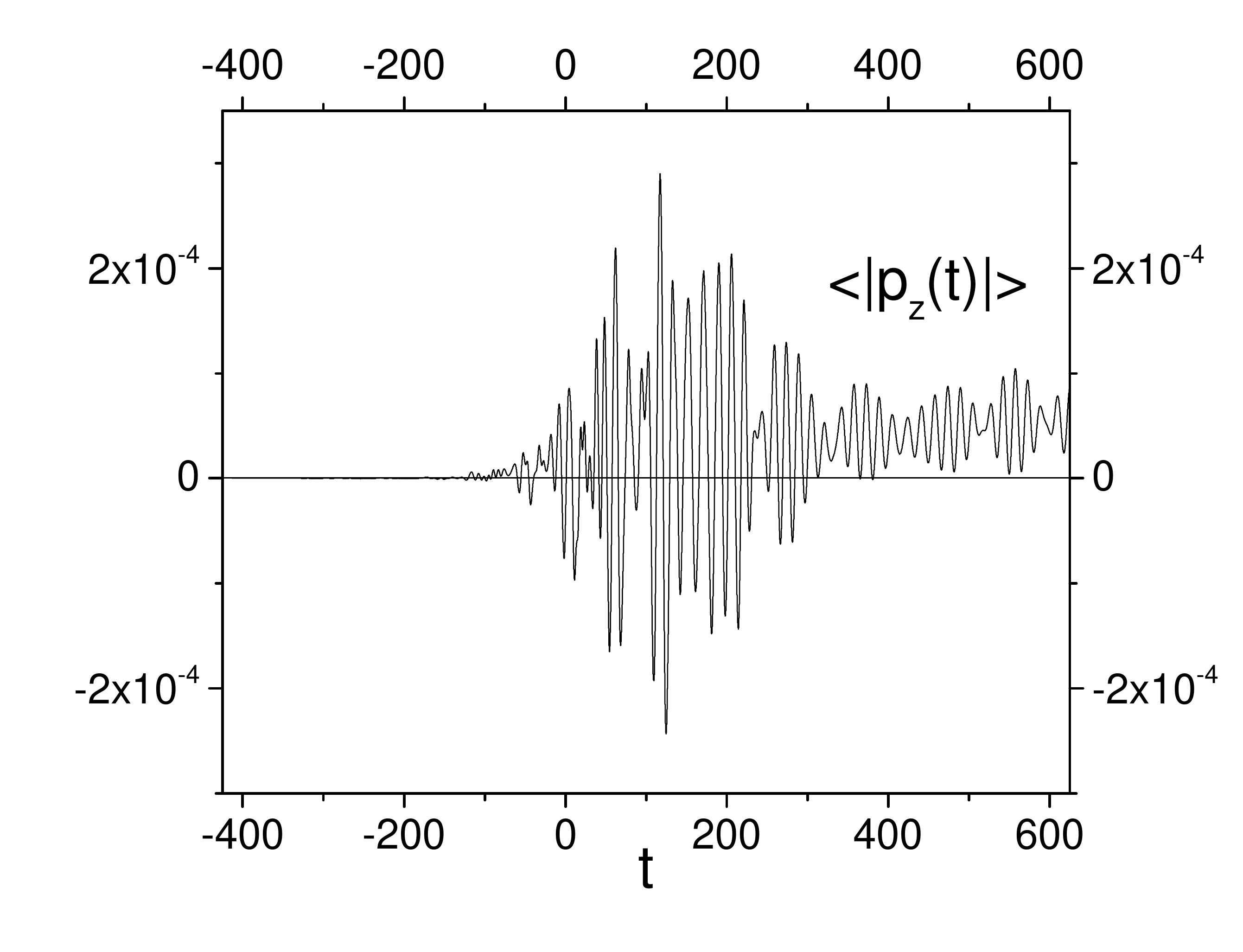}}
\end{subfigure}
\end{center}
\caption{The values $X(t)$,$Z(t)$, $P_x(t)$, $P_z(t)$, $\langle|p_x(t)|\rangle$ and $\langle|p_z(t)|\rangle$ calculated for the laser field with $I=10^{14}\frac{W}{cm^2}$ and $\lambda=800nm$ ($\omega=0.057$a.u.). The time-dependence of the laser pulse $E(\omega t)$ (\ref{field0}) is also presented at $z=0$. The time is given in a.u. ($\frac{\hbar}{m_e(\alpha c)^2} =2.42\times 10^{-17} sec$)}.
\label{fig:Fig1}
\end{figure}
Here and below, we do not present the calculated y-components, since they are several orders of magnitude smaller than the x- and z-components due to the absence of terms depending on y in the coupling potential ({\ref{interaction2}).

Next, we have calculated the spectral densities $|P_x(\omega)|^2$, $|P_z(\omega)|^2$ and $|p_x(\omega)|^2$, $|p_z(\omega)|^2$ of the kinetic energies of the atom and electron (see definitions ({\ref{ekinM}-\ref{ekint})), which are shown in Fig. 2 for the region of excited states of the atom $\hbar\omega \geq -\frac{1}{8}=-0.125$ (see the lower graph in Fig. 2). Here we should note the repetition of the shape of the x- component of the distribution curve $|p_x(\omega)|^2$ of the electron kinetic energy of in the spectral density $|P_x(\omega)|^2$ of the atom kinetic energy. The same effect is qualitatively repeated for the components $|p_z(\omega)|^2$ and $|P_z(\omega)|^2$. It also shows the calculated populations $W_n=\sum_{lm}|\langle \phi_{nlm}({\bi r})|\psi({\bi r},t=(n_T+1)T)\rangle|^2$ of the low-lying states of the hydrogen atom after the end of the action of the laser pulse.
\begin{figure}[!]
\begin{center}
\includegraphics[width=0.6\columnwidth]{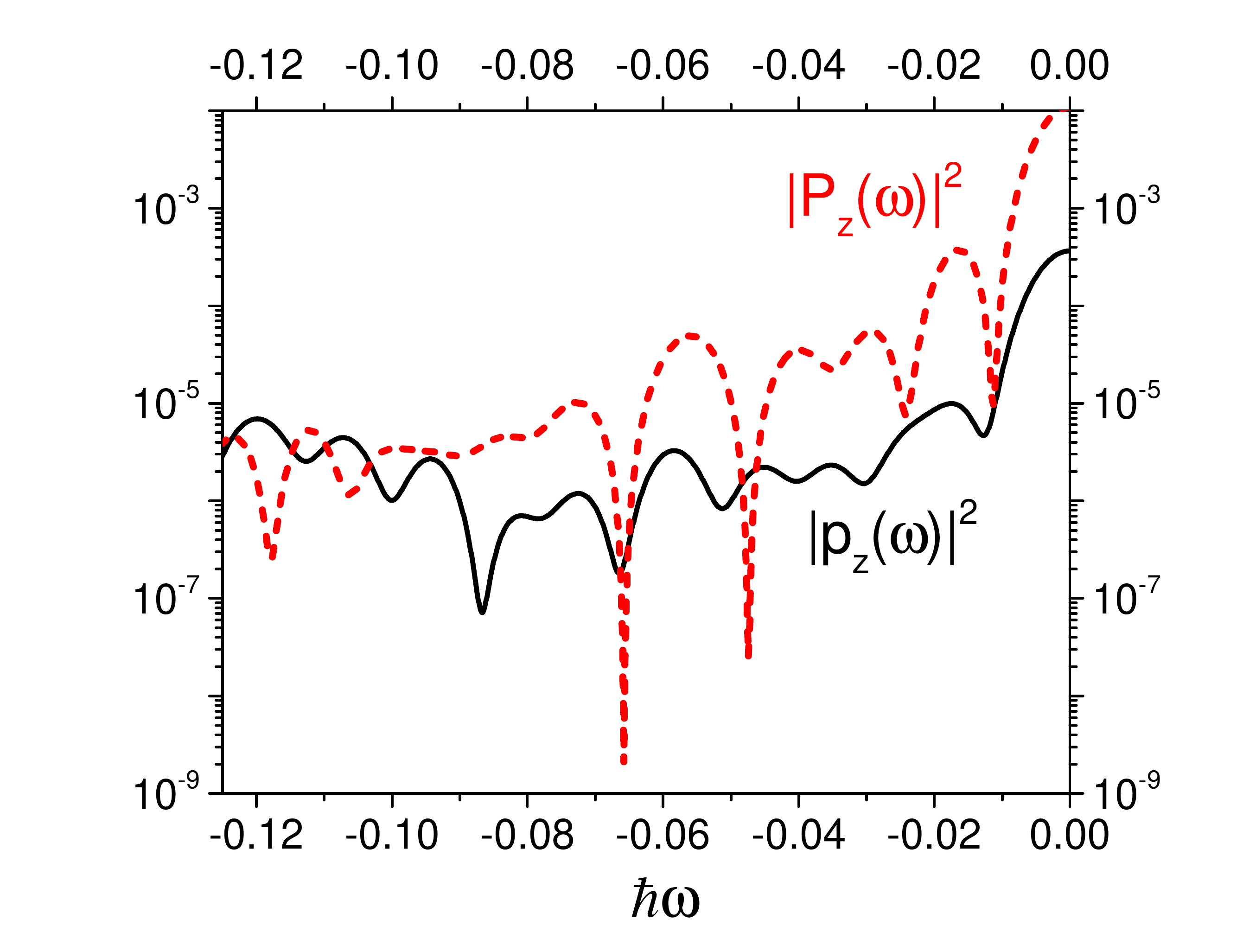}\\
\includegraphics[width=0.6\columnwidth]{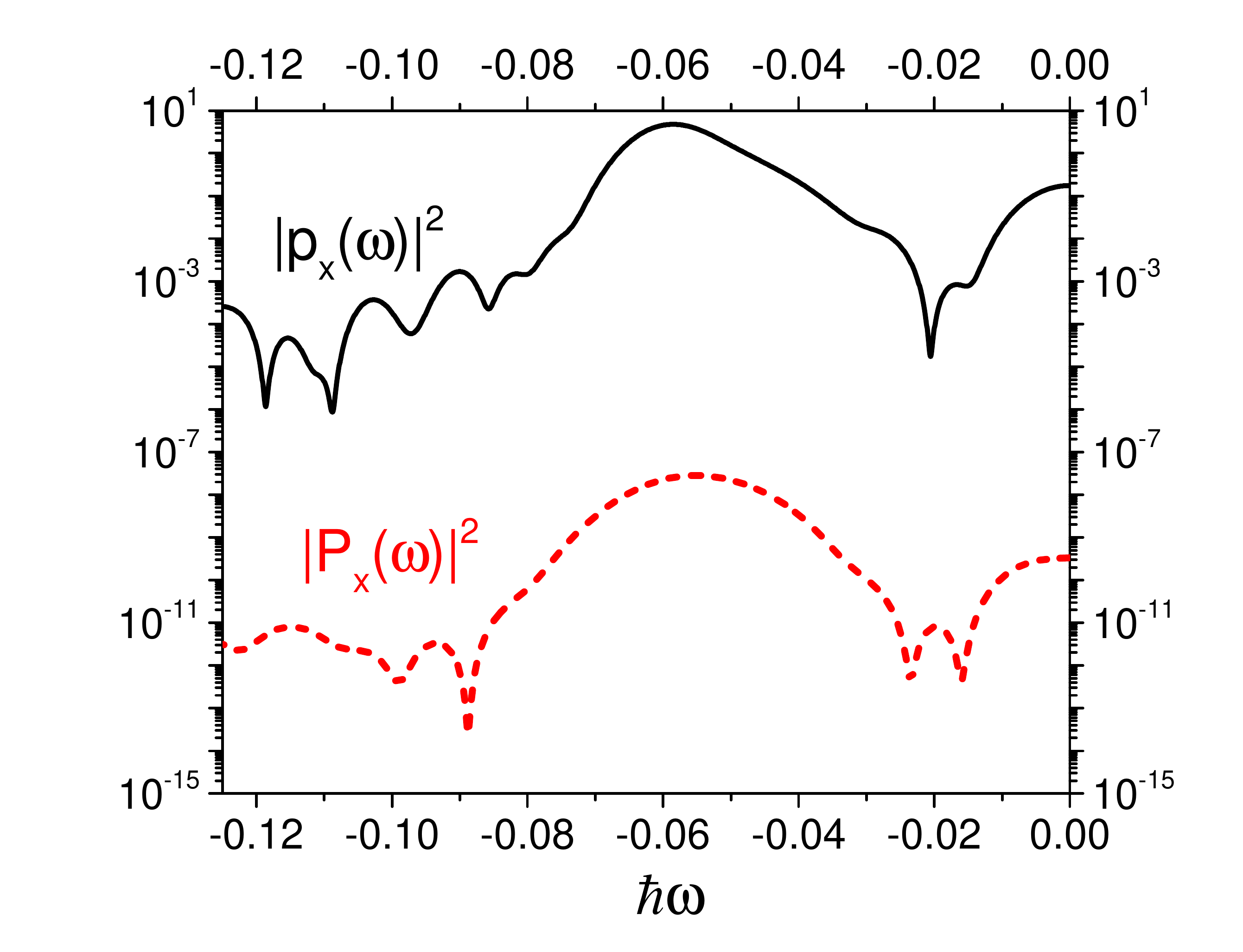}\\
\includegraphics[width=0.6\columnwidth]{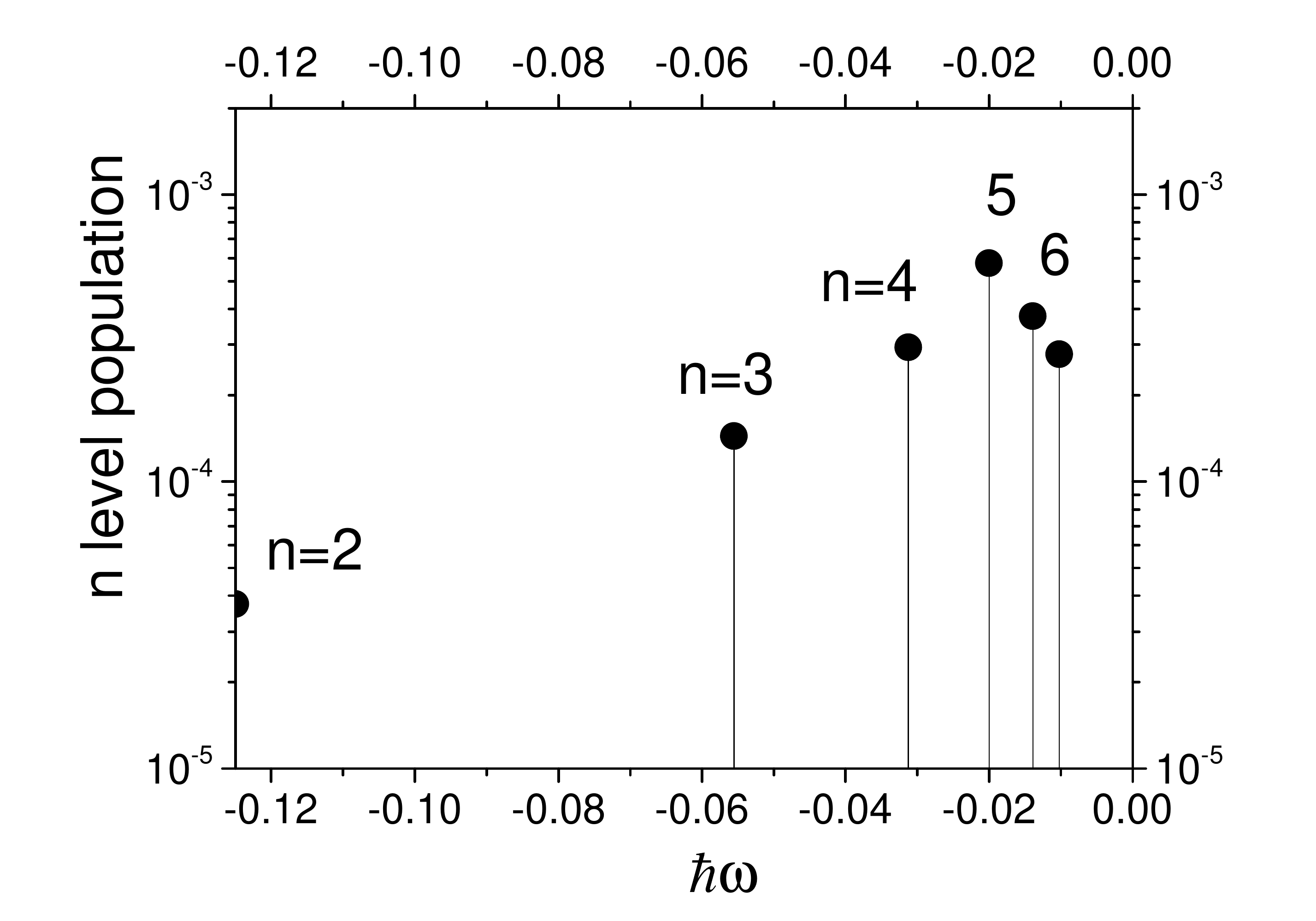}\\
\end{center}
\caption{The values $|P_x(\omega)|^2$, $|P_z(\omega)|^2$, $|p_x(\omega)|^2$,$|p_z(\omega)|^2$ and the populations $W_n$ calculated for the laser field with $I=10^{14}\frac{W}{cm^2}$ and $\lambda=800nm$ ($\omega=0.057$a.u.).}
\label{fig:Fig2}
\end{figure}

In Figs. 3 and 4, we present the results of calculations with doubling the laser radiation frequency $\lambda=400nm$ ($\omega=0.114$). As we can see in Fig. 3, this leads to an increase in the acceleration of the atom due to the amplification of the term (\ref{interaction2}) mixing the CM- motion and the relative motion of the electron in the atom. As in the previous case, at times approaching $t\sim 0$  higher harmonics arise corresponding to the transitions $n\rightarrow n'=2,3...$ in the quantities $\langle|p_x(t)|\rangle$ and $\langle|p_z(t)|\rangle$. As in the previous case, the calculated spectral densities $|P_s(\omega)|^2$ repeat the shape of the electron kinetic energy distributions $|p_s(\omega)|^2$ in the atom. Note also some shifts in energy between distributions $|P_s(\omega)|^2$ and $|p_s(\omega)|^2$, which is visible as well in Figs. 2 for $\lambda=800nm$. This effect can be explained by the fact that the spectral densities $|P_s(\omega)|^2$ for atoms are calculated in the laboratory frame and the electron distributions $|p_s(\omega)|^2$ in the accelerated CM- frame.
\begin{figure}[!]
\begin{center}
\begin{subfigure}{
\includegraphics[width=0.45\columnwidth]{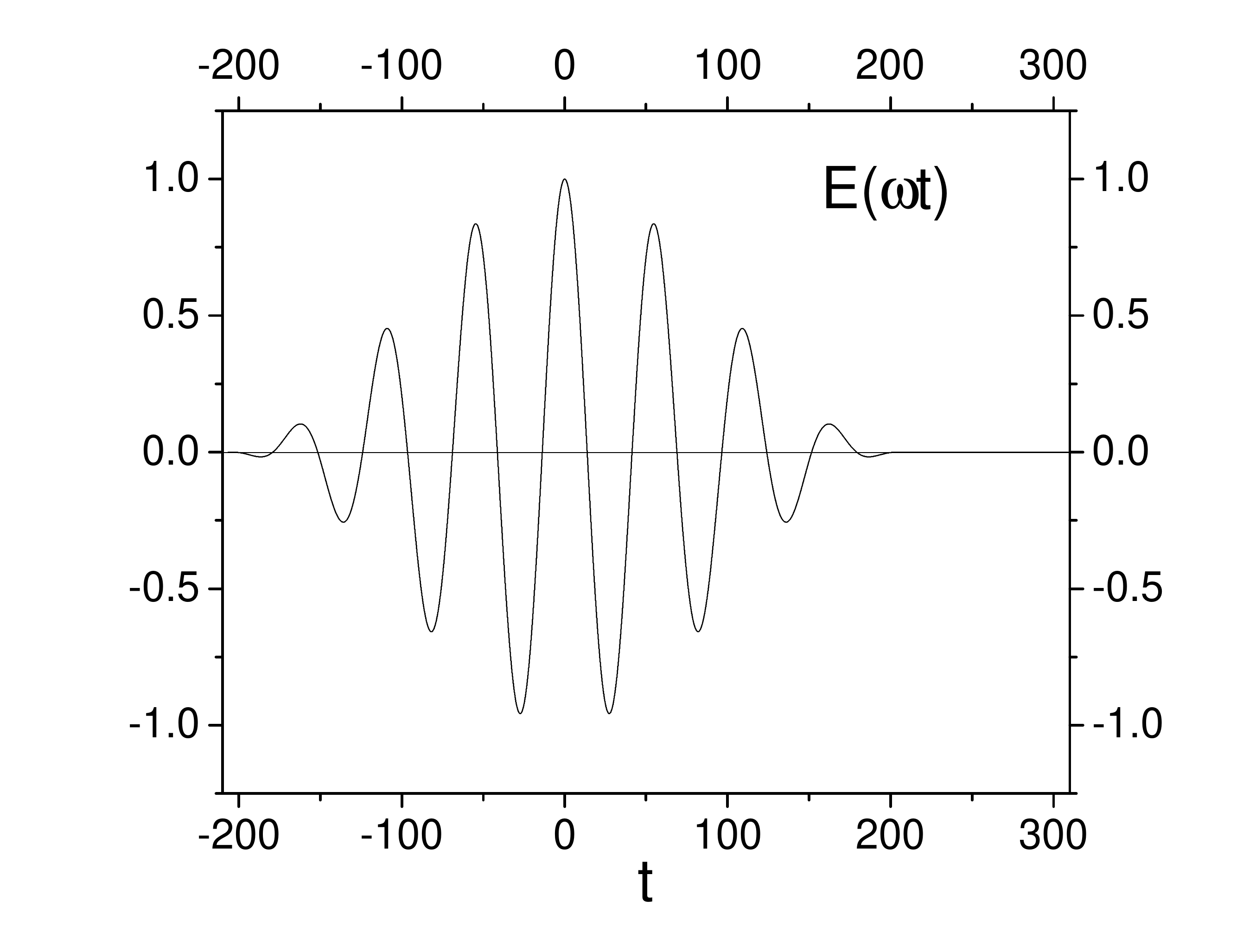} \includegraphics[width=0.45\columnwidth]{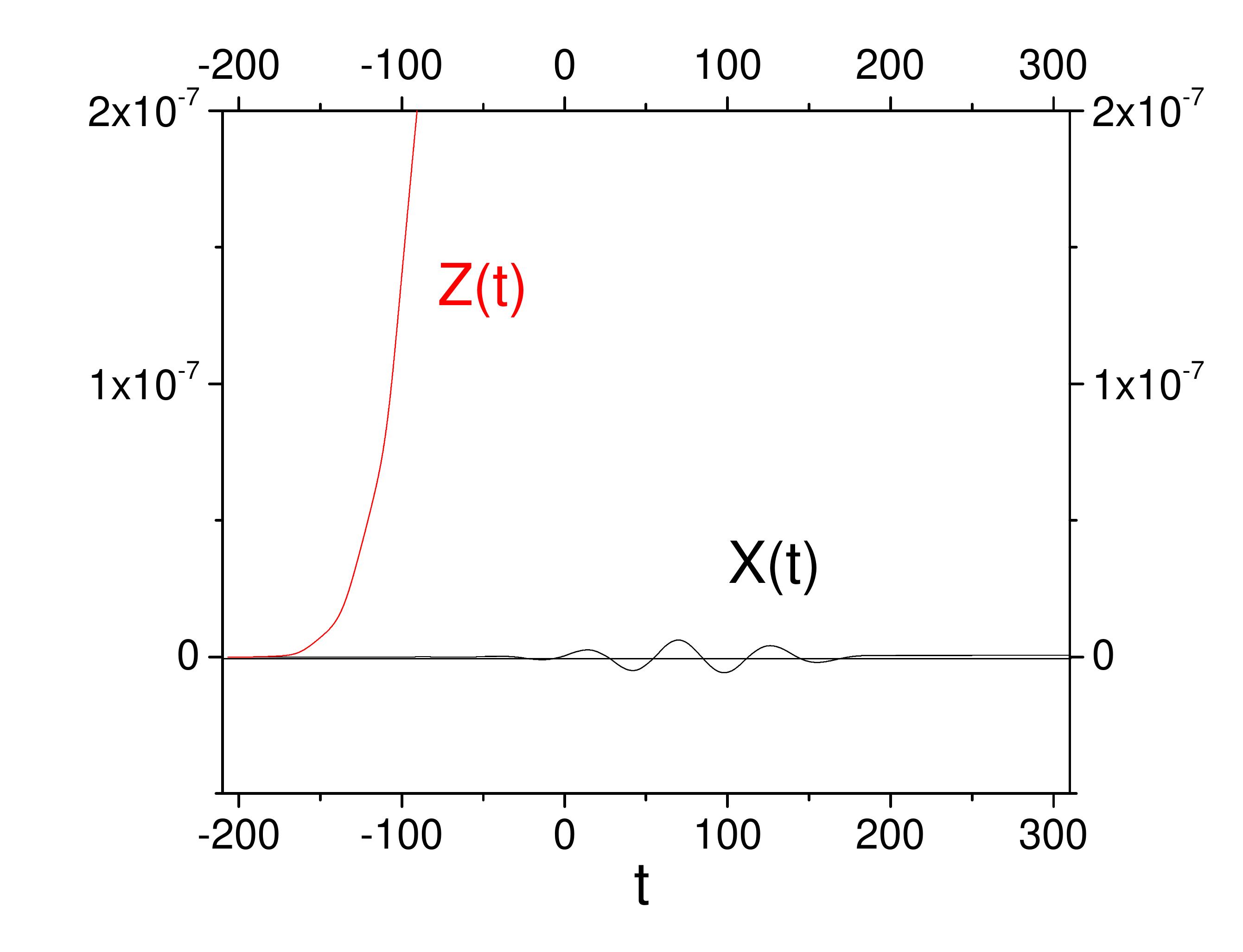}}
\end{subfigure}\\
\begin{subfigure}{
\includegraphics[width=0.45\columnwidth]{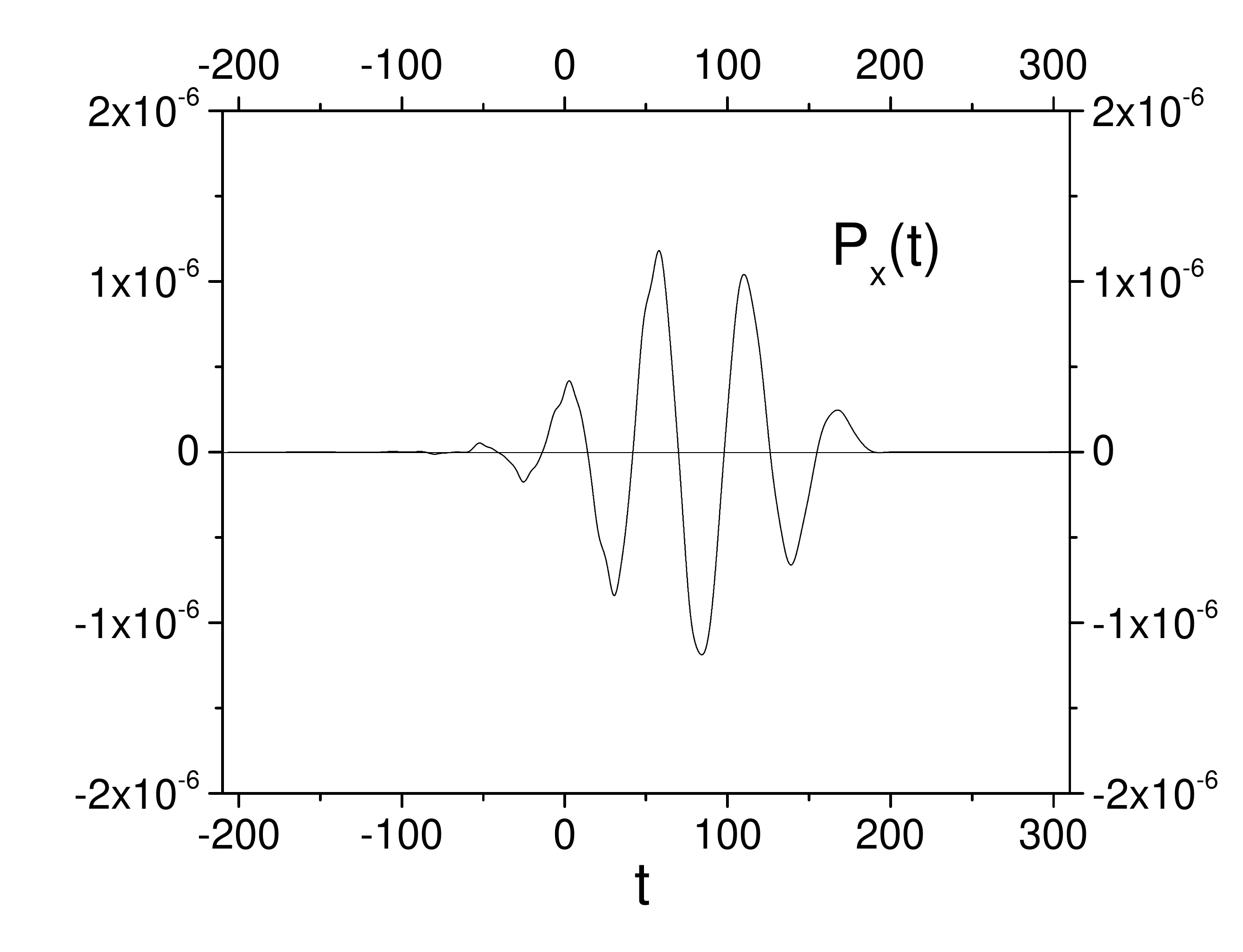} \includegraphics[width=0.45\columnwidth]{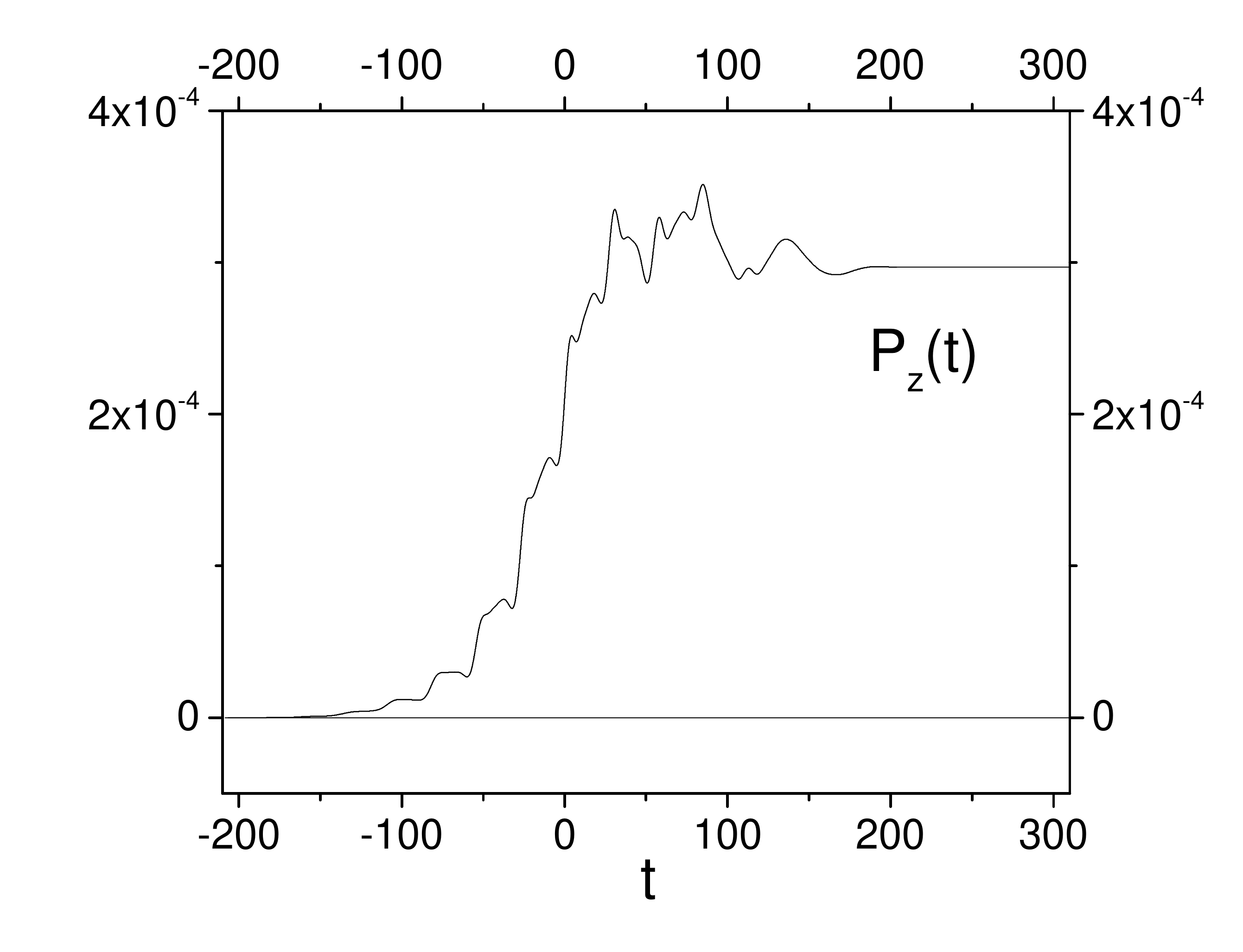}}
\end{subfigure}
\begin{subfigure}{
\includegraphics[width=0.45\columnwidth]{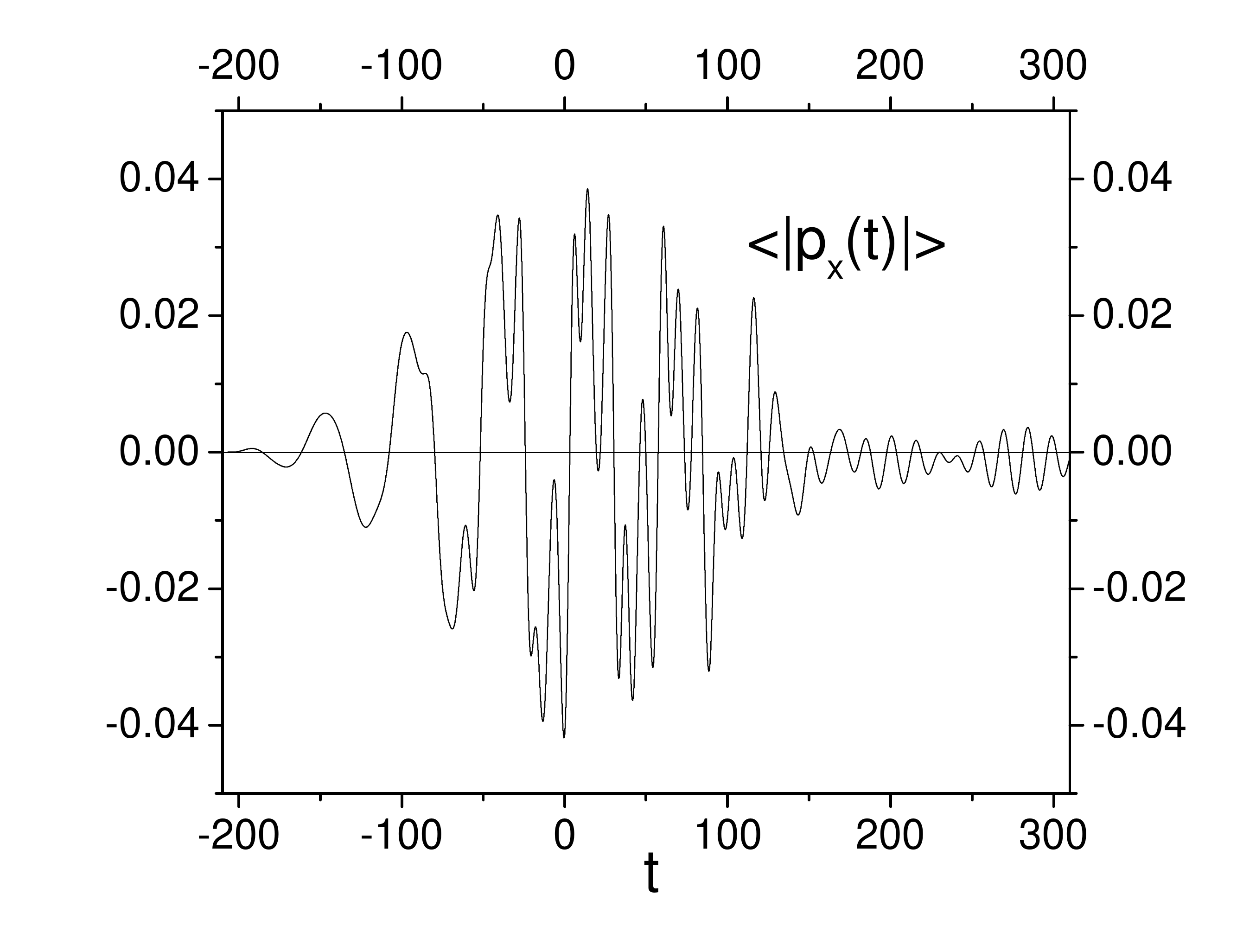} \includegraphics[width=0.45\columnwidth]{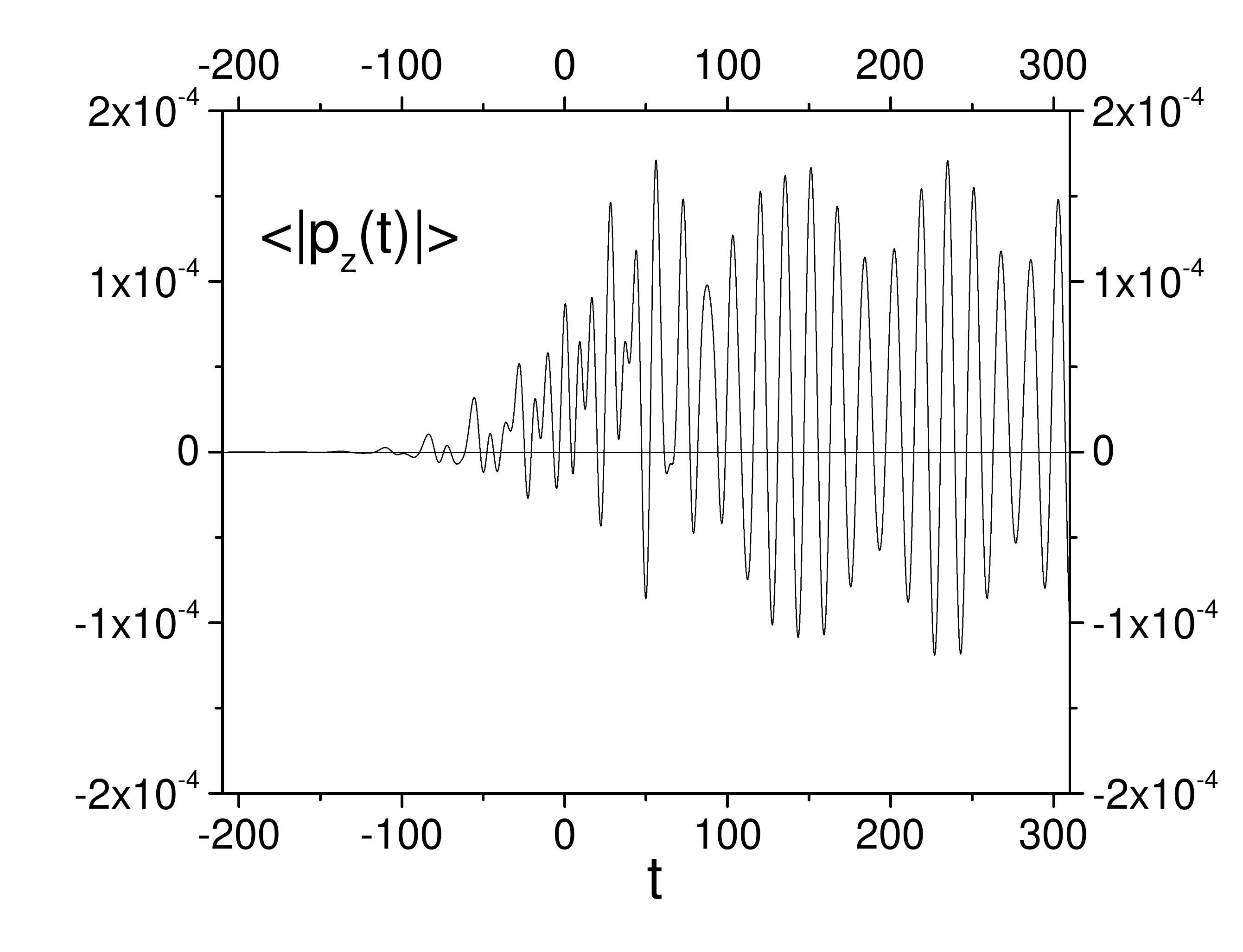}}
\end{subfigure}
\end{center}
\caption{The values $X(t)$,$Z(t)$, $P_x(t)$, $P_z(t)$, $\langle|p_x(t)|\rangle$ and $\langle|p_z(t)|\rangle$ calculated for the laser field with $I=10^{14}\frac{W}{cm^2}$ and $\lambda=400nm$ ($\omega=0.114$a.u.). The time-dependence of the laser pulse $E(\omega t)$ (\ref{field0}) is also presented at $z=0$.}
\label{fig:Fig3}
\end{figure}
\begin{figure}[!]
\begin{center}
\includegraphics[width=0.6\columnwidth]{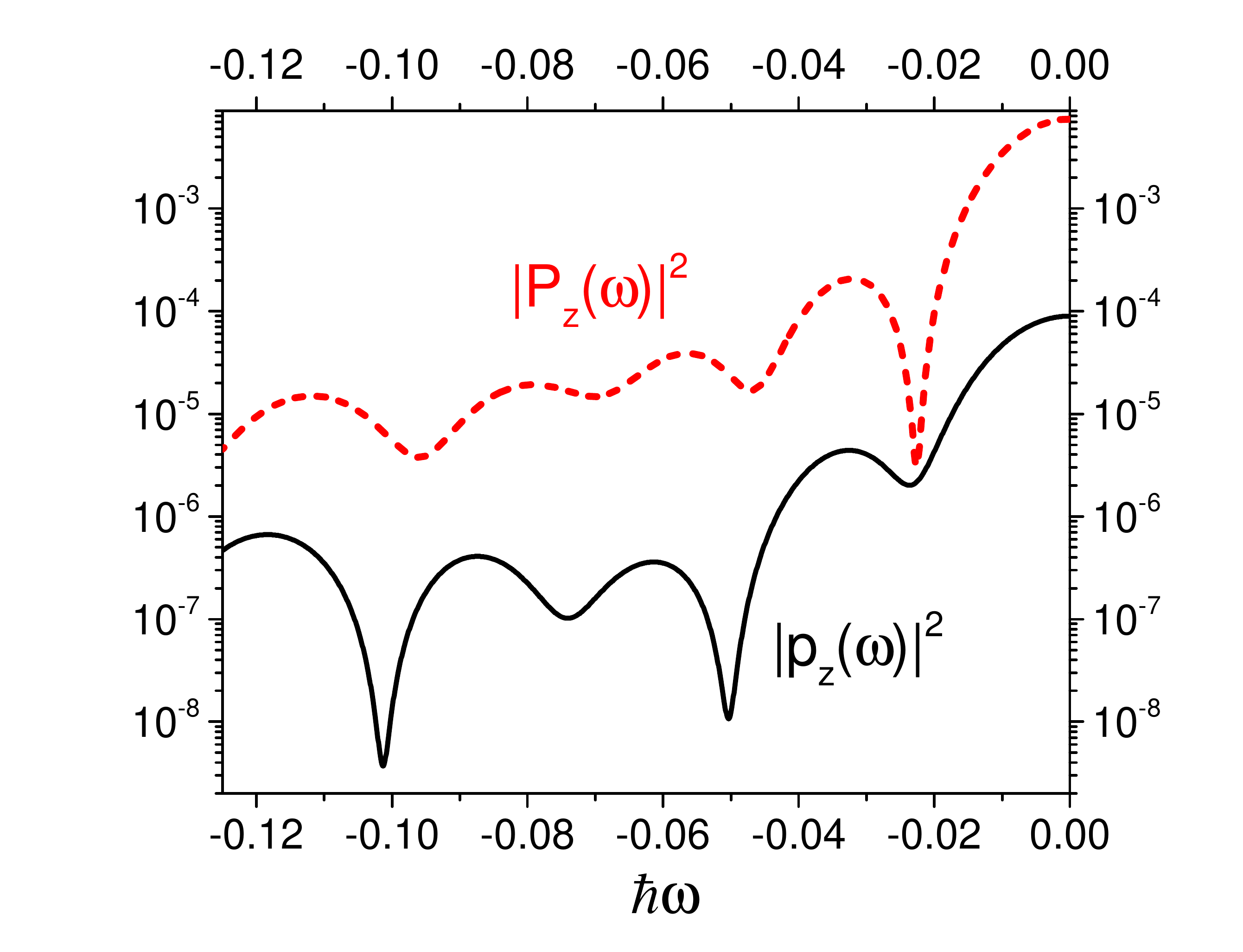}\\
\includegraphics[width=0.6\columnwidth]{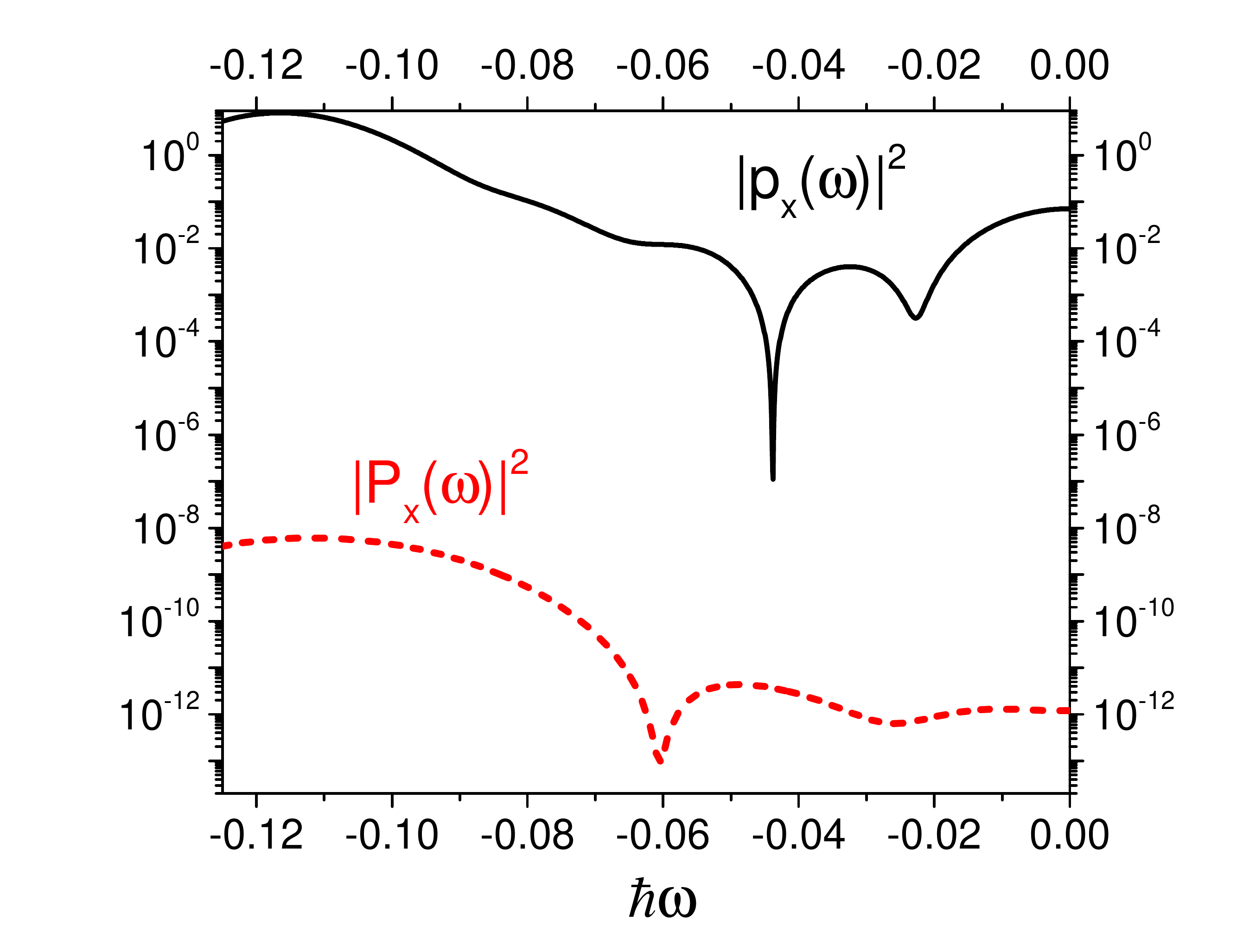}\\
\includegraphics[width=0.6\columnwidth]{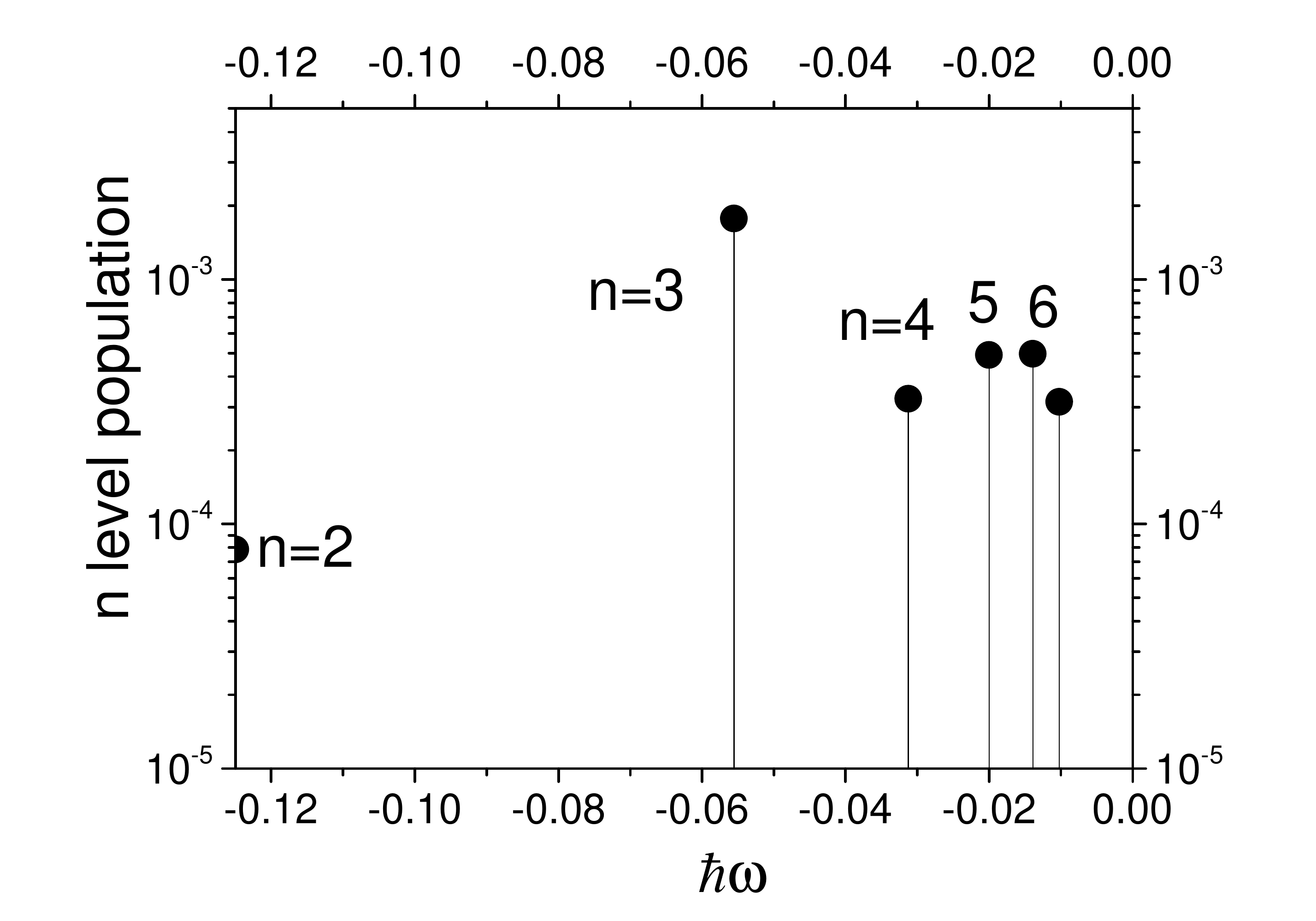}\\
\end{center}
\caption{The values $|P_x(\omega)|^2$, $|P_z(\omega)|^2$, $|p_x(\omega)|^2$,$|p_z(\omega)|^2$ and the populations $W_n$ calculated for the laser field with $I=10^{14}\frac{W}{cm^2}$ and $\lambda=400nm$ ($\omega=0.5$a.u.).}
\label{fig:Fig4}
\end{figure}

Finally, we have performed calculations for $\lambda=90nm$ ($\omega=0.5$) that gives the strongest coupling between the CM- and relative electron motions. The results of these calculations are presented in Figs. 5 and 6. Amplification of the coupling term (\ref{interaction2}) in the Hamiltonian of the problem (\ref{hamiltonian}) due to an increase in $\omega$ leads to even greater acceleration of the atom in the direction of the laser pulse propagation compared to the previous cases. The amplitude of the atomic momentum oscillations in the direction of polarization during the interaction with the laser field exceeds the previous values. However, after the laser is turned off, the atomic momentum in this direction remains negligibly small, as before the interaction with the laser pulse. Since the frequency of transitions $n=1 \rightarrow n'=2,3,...$ between atomic levels in this case is less than the frequency of laser radiation $\Omega \sim \frac{1}{2}-\frac{1}{8} < \omega=0.5$, high-frequency components are not observed in $\langle|p_x(t)|\rangle$ and $\langle|p_z(t)|\rangle$. Moreover, at long times $t\geq 20$ these transitions lead to some slowing down of the $\langle|p_x(t)|\rangle$ oscillations. Figure 6 shows the calculated spectral densities $|P_s(\omega)|^2$ and $|p_s(\omega)|^2$. As in the previous cases, the shapes of the spectral density curves of the atomic kinetic energy repeat the shapes of the electron momentum distributions. It should be noted here that we found no dependence on the frequency in the distributions $|P_x(\omega)|^2$ and $|p_x(\omega)|^2$.
\begin{figure}[!]
\begin{center}
\begin{subfigure}{
\includegraphics[width=0.45\columnwidth]{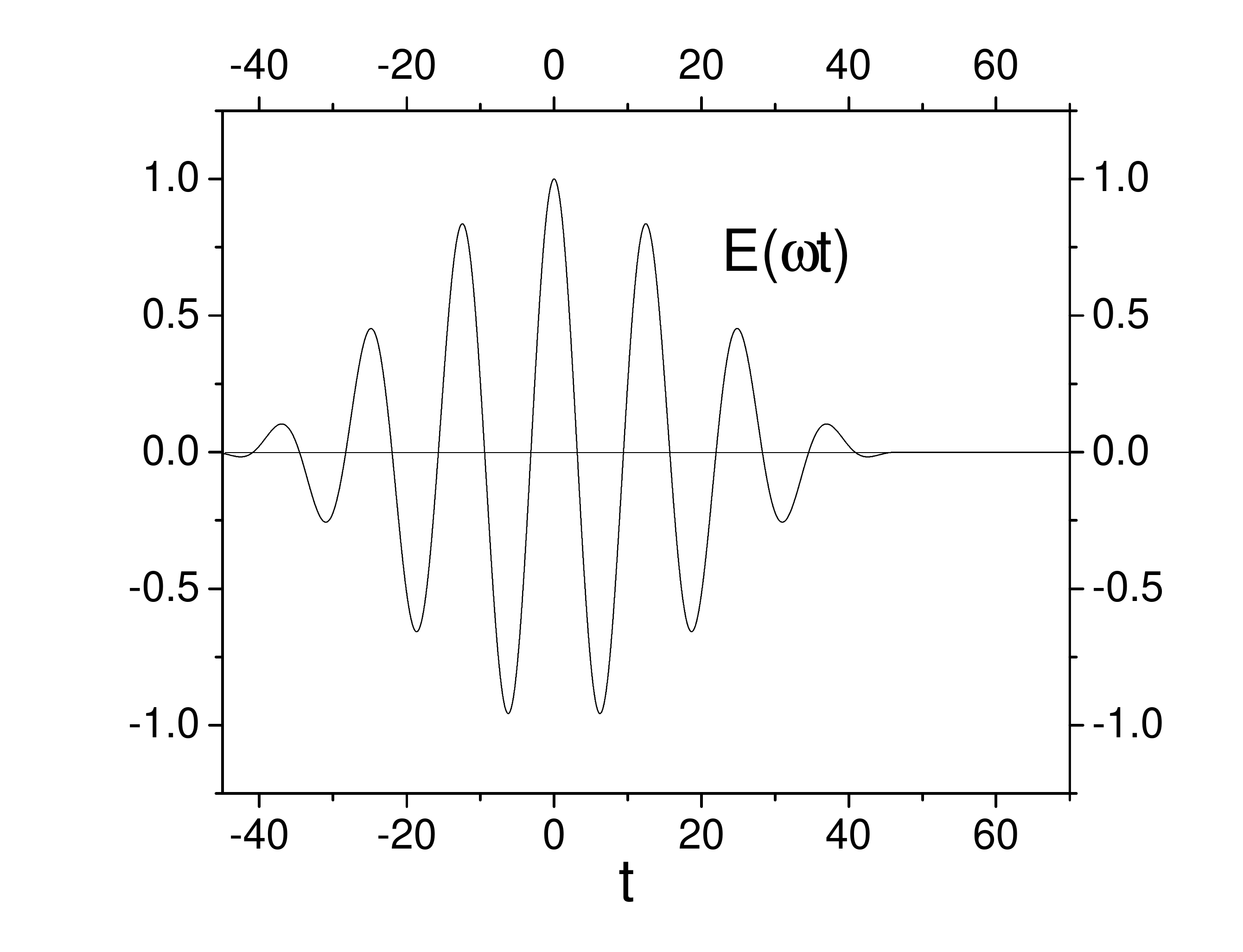} \includegraphics[width=0.45\columnwidth]{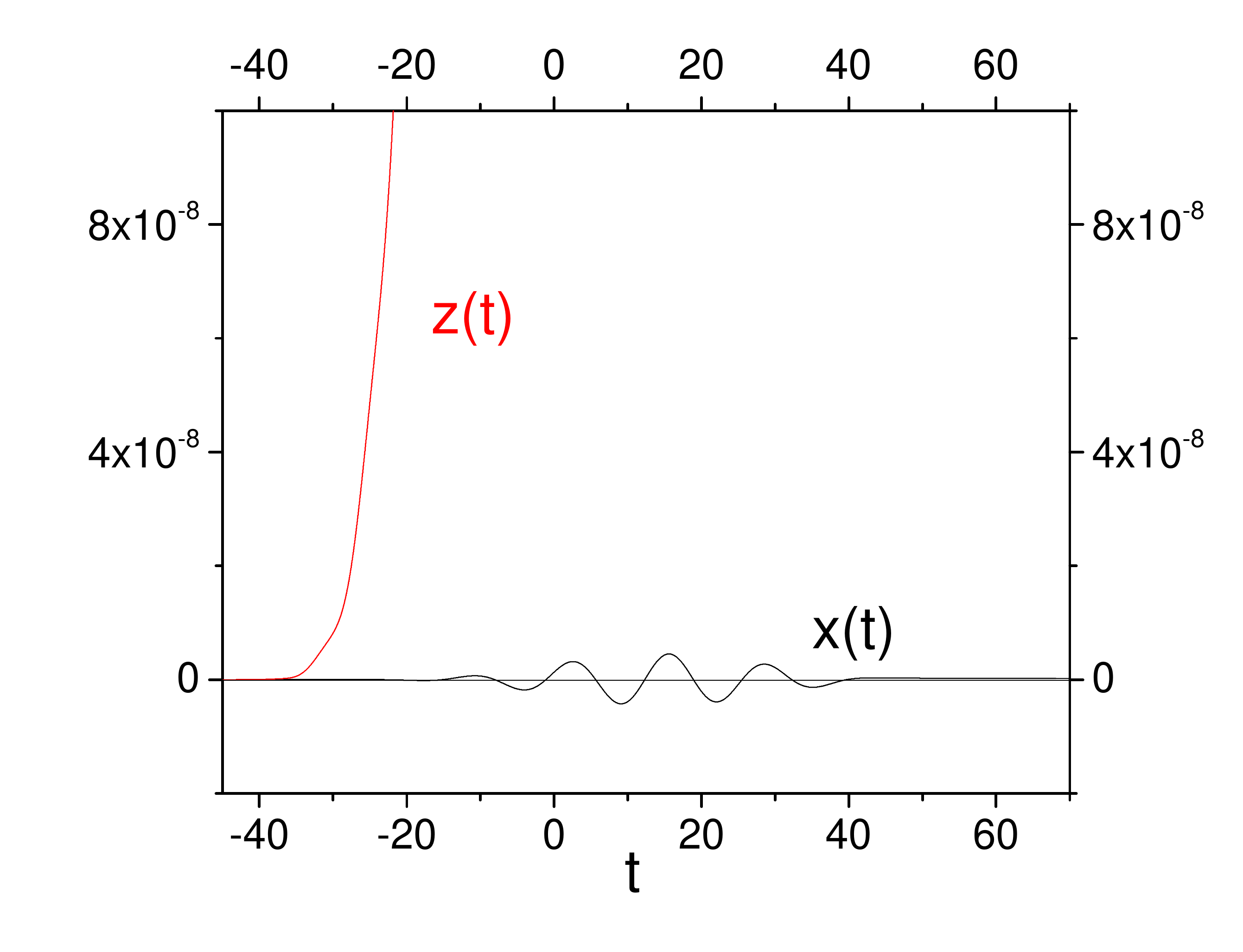}}
\end{subfigure}\\
\begin{subfigure}{
\includegraphics[width=0.45\columnwidth]{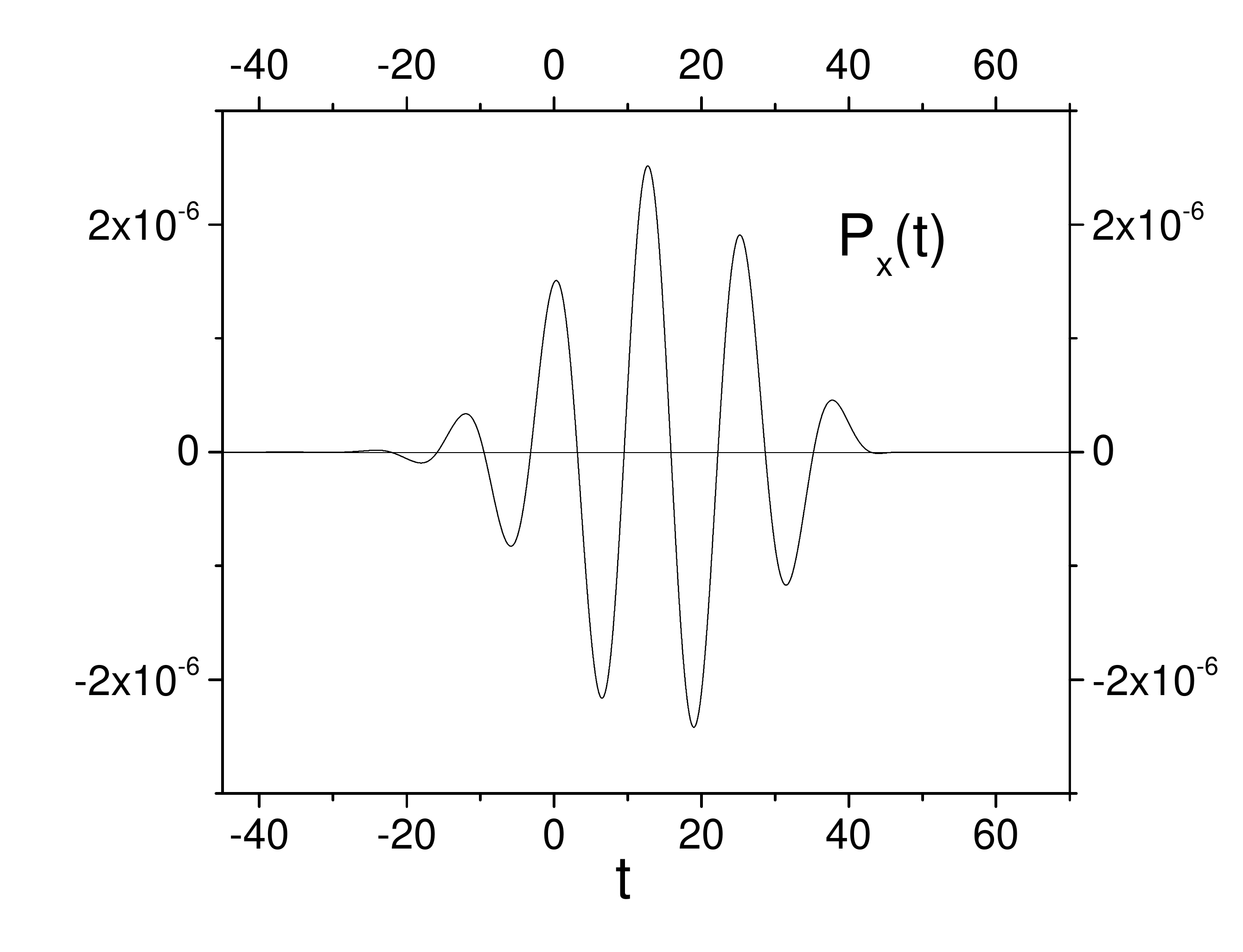} \includegraphics[width=0.45\columnwidth]{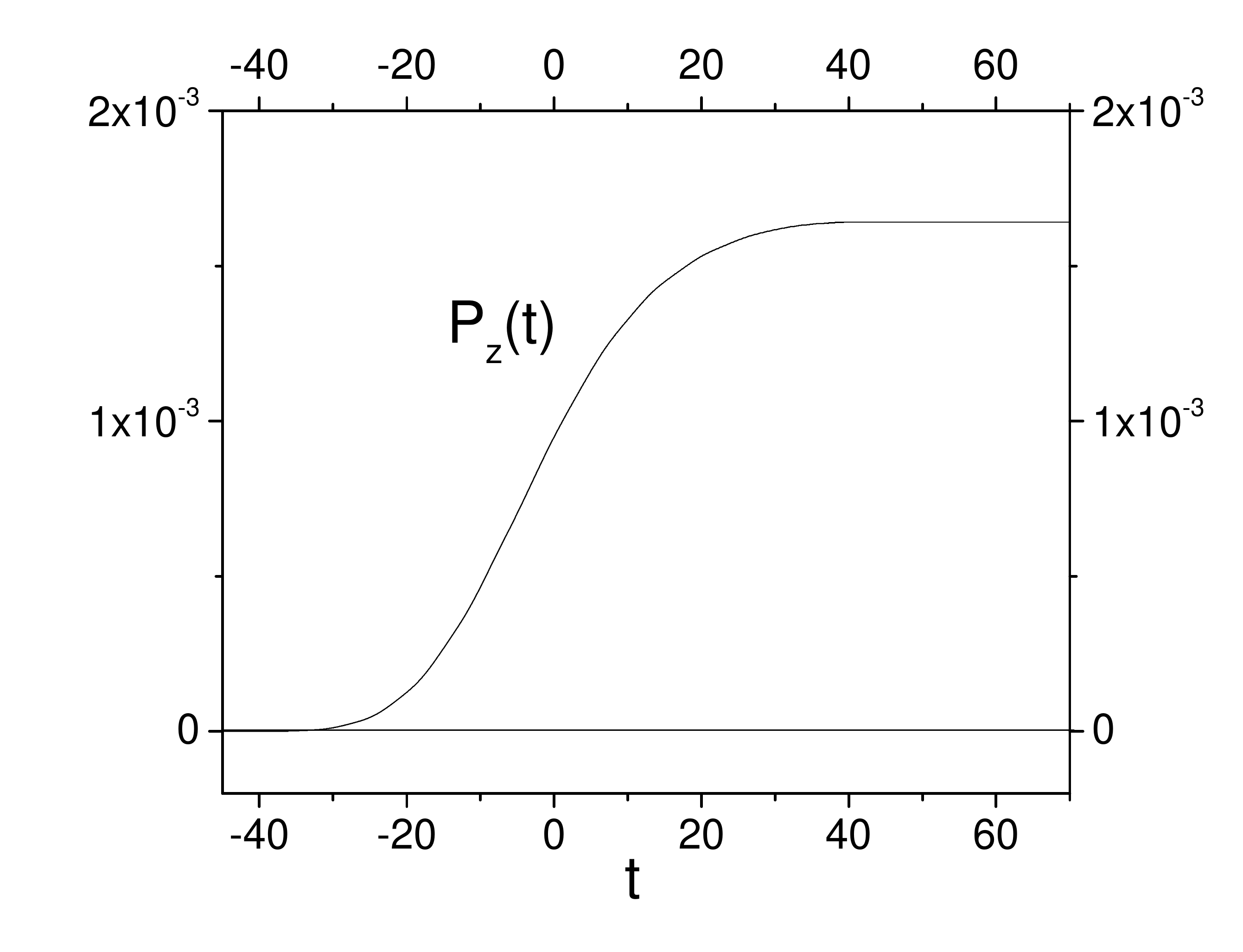}}
\end{subfigure}
\begin{subfigure}{
\includegraphics[width=0.45\columnwidth]{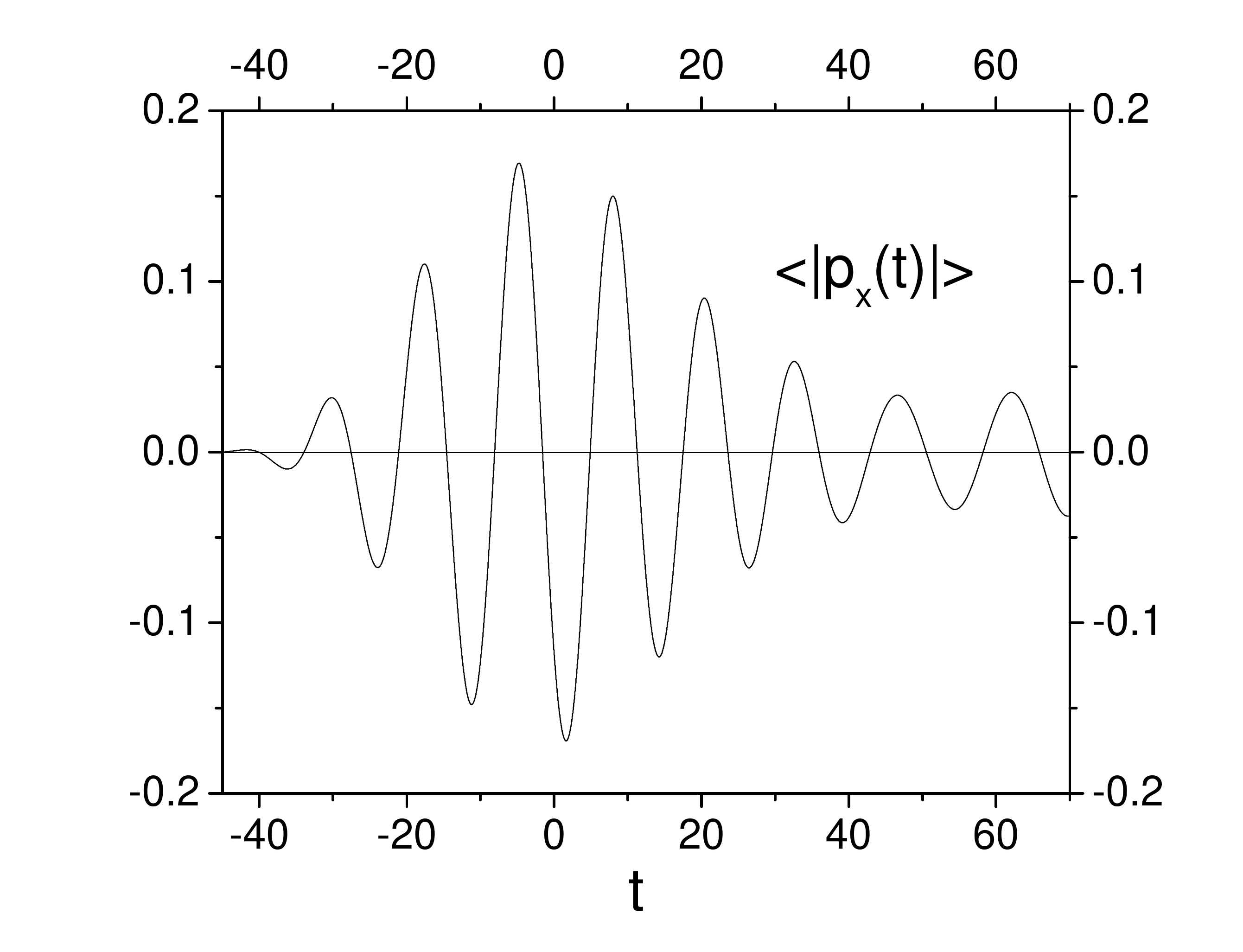} \includegraphics[width=0.45\columnwidth]{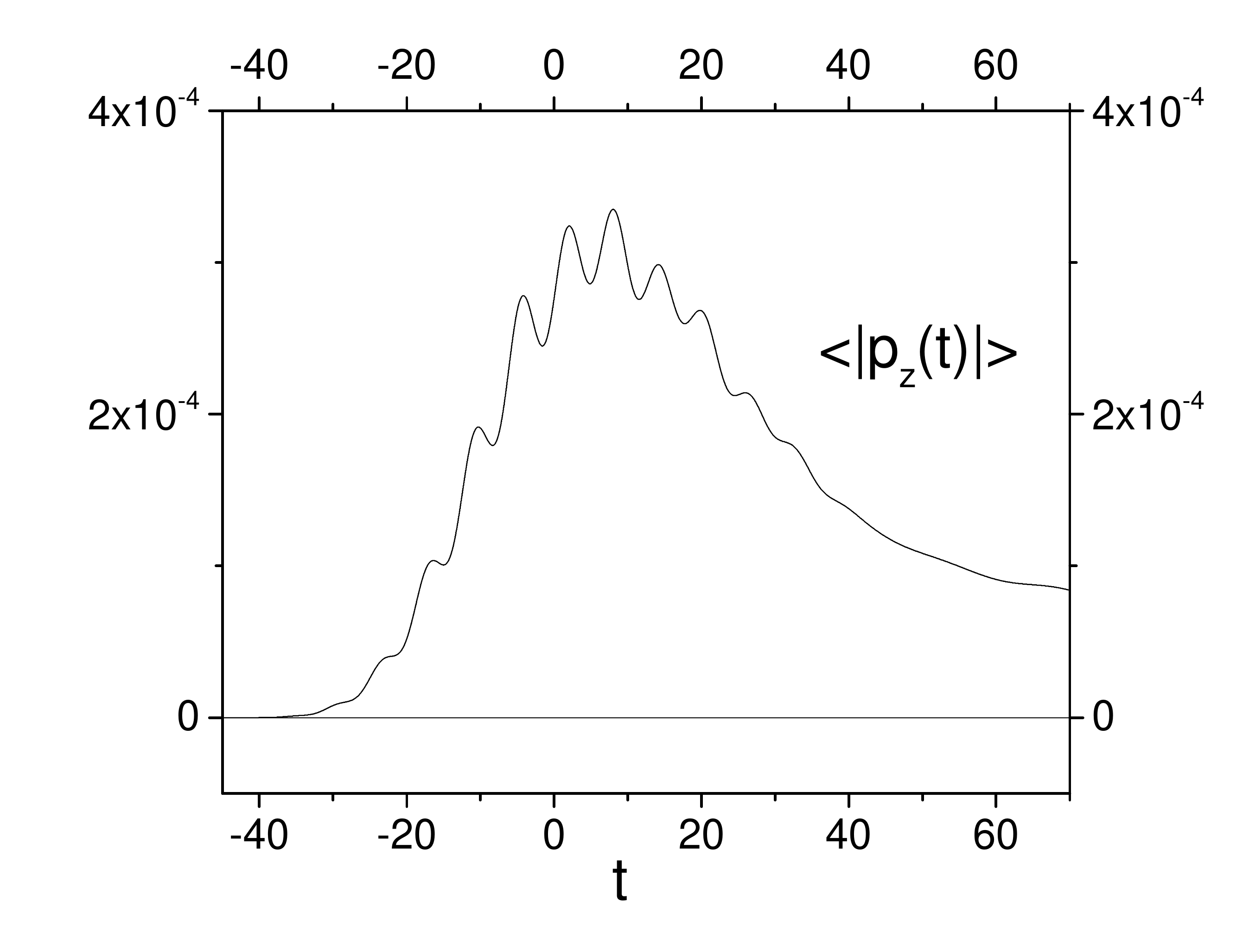}}
\end{subfigure}
\end{center}
\caption{The values $X(t)$,$Z(t)$, $P_x(t)$, $P_z(t)$, $\langle|p_x(t)|\rangle$ and $\langle|p_z(t)|\rangle$ calculated for the laser field with $I=10^{14}\frac{W}{cm^2}$ and $\lambda=90nm$ ($\omega=0.5$a.u.). The time-dependence of the laser pulse $E(\omega t)$ (\ref{field0}) is also presented at $z=0$.}
\label{fig:Fig5}
\end{figure}
\begin{figure}[!]
\begin{center}
\includegraphics[width=0.6\columnwidth]{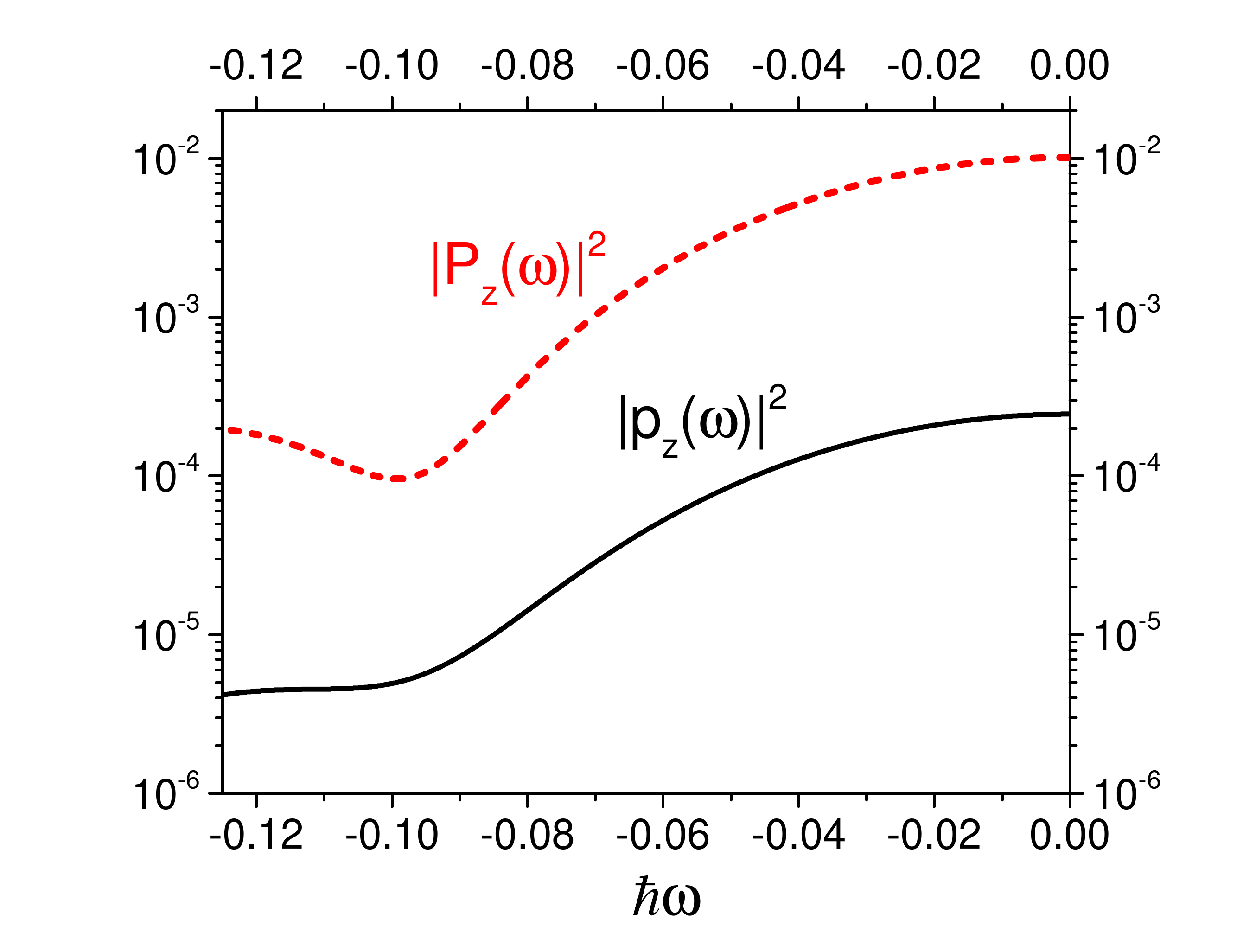}\\
\includegraphics[width=0.6\columnwidth]{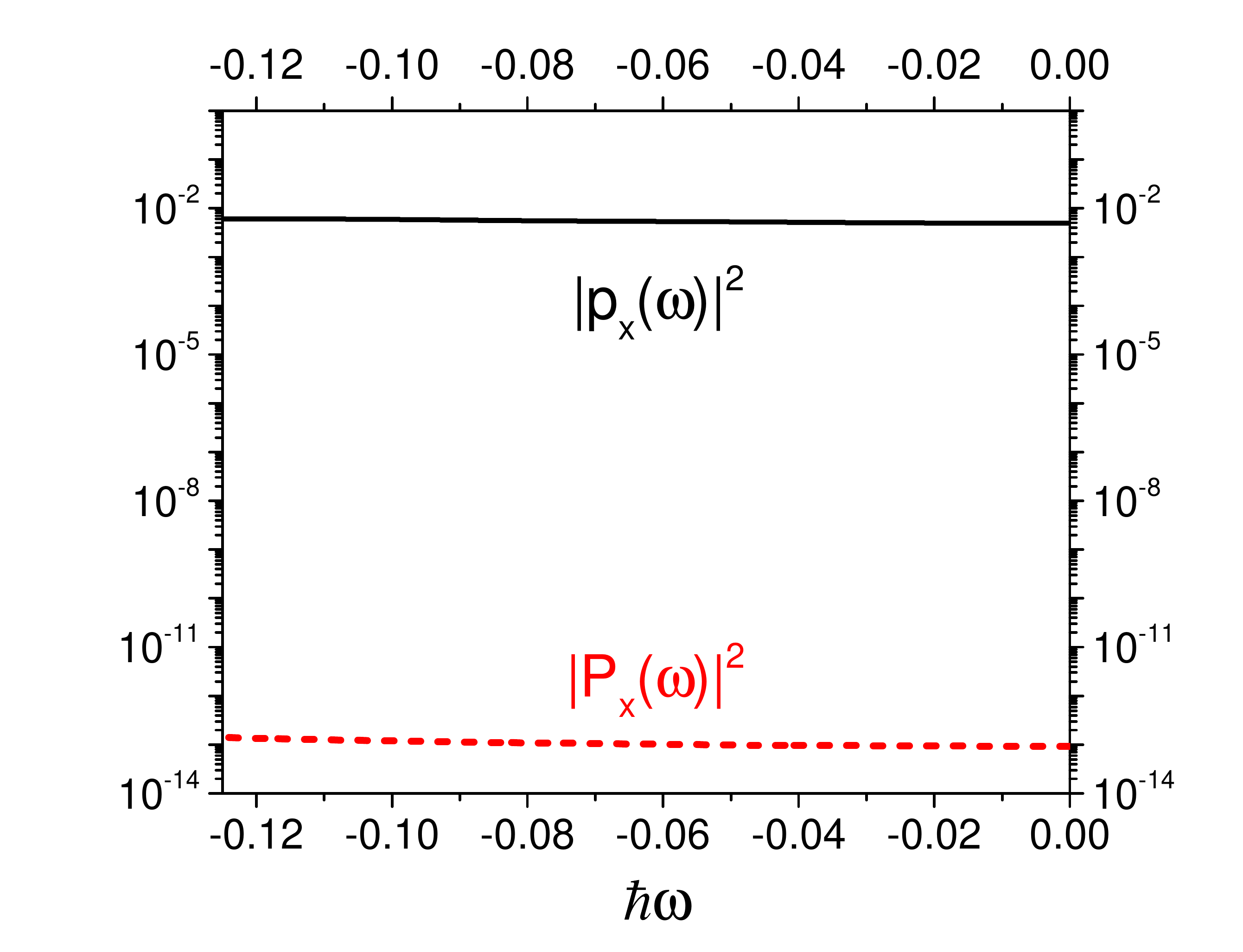}\\
\includegraphics[width=0.6\columnwidth]{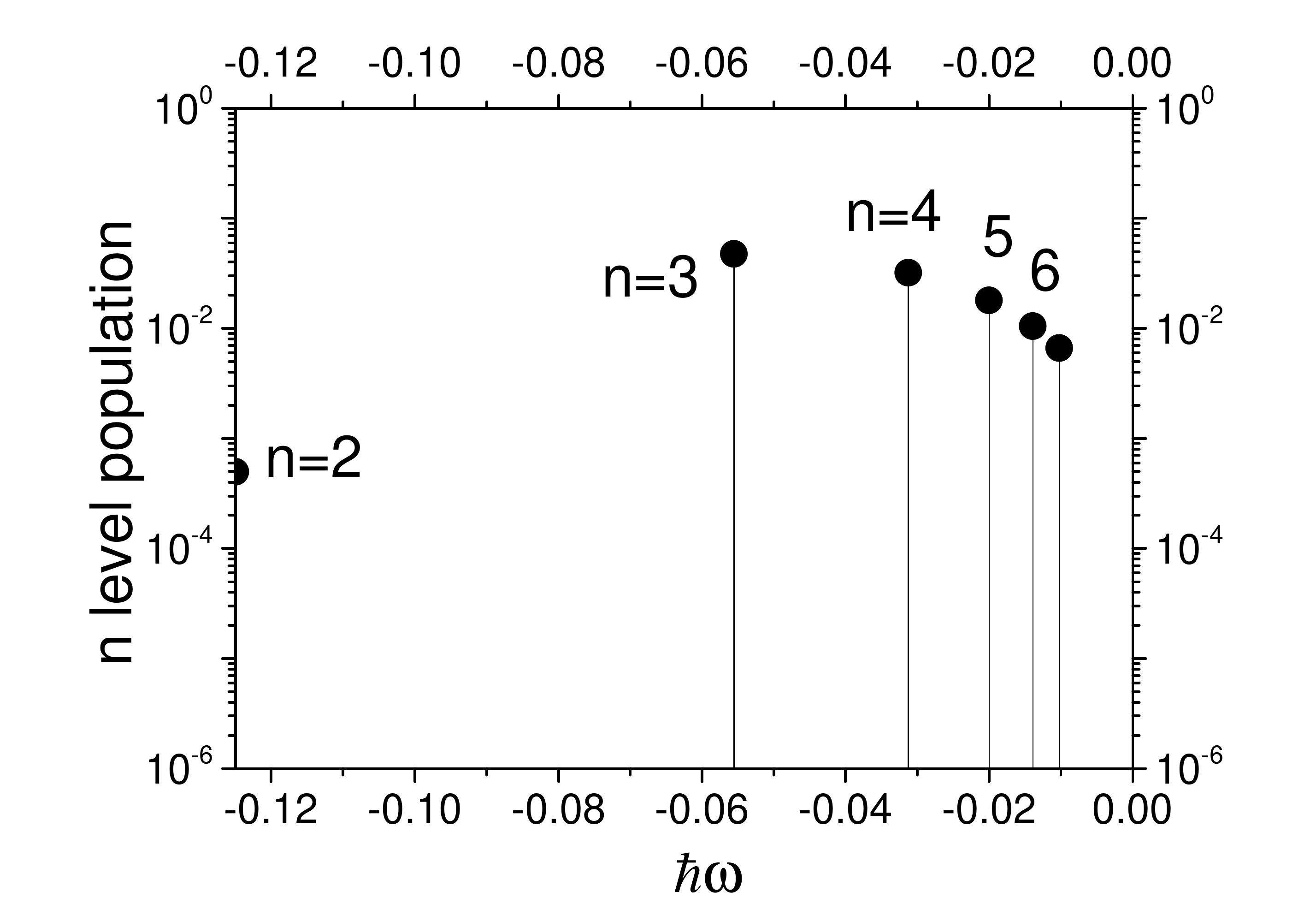}\\
\end{center}
\caption{The values $|P_x(\omega)|^2$, $|P_z(\omega)|^2$, $|p_x(\omega)|^2$,$|p_z(\omega)|^2$ and the populations $W_n$ calculated for the laser field with $I=10^{14}\frac{W}{cm^2}$ and $\lambda=90nm$ ($\omega=0.5$a.u.).}
\label{fig:Fig6}
\end{figure}

Here we must emphasize the following fact. In Figs. 2, 4, and 6, we present the calculated curves $|P_s(\omega)||^2$ in the region $-0.125 \leq\omega < 0$. However, the Maxwellian peak at room temperature corresponds to $300K \simeq 0.04eV \simeq \hbar\omega\simeq 10^{-3}a.u.$ very close to the point $\hbar\omega\sim 0$, i.e. practically out of the region of the potentially supposed measurements of the spectral densities $|P_s(\omega)|^2$ where we performed calculations.


\section{Conclusion}\label{Conclusion}
The paper describes a quantum-quasiclassical approach to the quantitative analysis of the 6D hydrogen atom in a strong laser field. This approach allowed us to study this problem outside the commonly used dipole approximation for the interaction of a laser field with an atom, when the electron variables cannot be separated from the CM variables. Calculations performed for different frequencies of laser radiation showed that an increase in frequency, which leads to an increase in the term coupling CM- and relative variables in the Hamiltonian of the problem, leads to an increase in the acceleration of an atom by a laser field.

We have found that the experimentally measured quantity, the spectral density of the atomic kinetic energy $|P_s(\omega)|^2$,  qualitatively repeats the shape of the distribution curve $|p_s(\omega)|^2$ of the kinetic energy of an electron in an atom. That is, it is shown that one can detect the quantum dynamics of an electron by measuring the distribution of the atom kinetic energy. Our result confirms the suggestion to apply the CM- velocity spectroscopy (a ``built-in'' classical instrument) to detect the internal quantum dynamics of an electron. The idea of using the CM- velocity spectroscopy for detecting electron dynamics was recently suggested in work~\cite{Patchkovskii} where, however, the correlation between the CM-velocity and the populations of atomic levels due to interaction with laser fields was analyzed in the simplified 3D quantum model for the hydrogen atom in laser fields. Here we analyzed and justified this idea in a more natural way by applying our quantum-quasiclassical approach for the 6D atom in a laser field, in which we simultaneously integrated the Schr\"odinger equation for the electron and the classical Hamiltonian equations for CM-variables. In this regard, it is interesting to note that recently the closeness of the momentum distributions of nuclei and electrons was observed during the ionization of a helium atom by the Compton scattering method~\cite{Kircher}. The role of nuclear motion in some problems of atomic and molecular physics was also discussed in the review article~\cite{Neudachin}.

The approach we have developed opens up the possibility of investigating the influence of CM motion in the non-dipole approximation on the ionization and excitation of atoms by strong laser fields, the radiation pressure in this problem, the generation of high harmonics, and other effects.

\ack
The author is grateful to S. Shadmehri for valuable discussions and help. The work was supported by the Russian Science Foundation, Grants No. 20-11-20257.

\appendix
\label{sec:appendix}

\section{Non-dipole interaction of hydrogen atom with laser field}

A particle with charge q and mass m in an electromagnetic field (1) is affected by the force
\begin{equation}
\label{eq:a1}
{\bf F}= q{\bf E}+\frac{q}{m}[{\bf p}\times{\bf B}]=qE(\omega t)\{{\bf n}_x+\frac{1}{mc}(\hat{p}_x{\bf n}_z-\hat{p}_z{\bf n}_x)\}\,.
\end{equation}
Here, we have neglected the term
\begin{equation}
-i\frac{q}{mc}E_0f(t)\frac{\partial}{\partial z}\cos(\omega t-kz) =-i\frac{q\omega}{mc^2}E_0f(t)\sin(\omega t-kz) \sim \frac{1}{c^2} \,\,\,
\end{equation}
of a higher order of smallness compared to $c^{-1}=137^{-1}$. Since the common factor $E(\omega t)=E_0f(t)\cos(\omega t-kz)$ in (A.1) also contains the parameter of smallness $\sim c^{-1}$ in term $kz=\omega c^{-1}z$, using the relation $\cos(\omega t-kz)\approx \cos(\omega t)-\omega c^{-1} z \sin(\omega t)$ whose accuracy is of the order of $c^{-1}$ and neglecting terms of higher orders than $\sim c^{-1}$, we reduce (A.1) to the form
\begin{equation}
{\bf F}= qE_0f(t)\{\cos(\omega t)[{\bf n}_x+\frac{1}{mc}(\hat{p}_x{\bf n}_z-\hat{p}_z{\bf n}_x)]+\frac{\omega}{c}\sin(\omega t)z{\bf n}_x\}\,.
\end{equation}
Further, using the well-known relation ${\bf F}({\bf r}) =-\nabla U({\bf r})$ connecting the vector field with a scalar potential field, we obtain the interaction potential of a charged particle with an electromagnetic field (1) up to terms $\sim c^{-1}$ inclusive
\begin{equation}
U({\bf r})=qE_0f(t)\{\cos(\omega t)x + \frac{1}{mc}\cos(\omega t)(z\hat{p}_x-x\hat{p}_z)+\frac{\omega}{c}\sin(\omega t)(xz)\}\,.
\end{equation}

We represent the interaction potential of a hydrogen atom with a laser field (1) as a sum of potentials (of type (A.4))
\begin{equation}
V({\bf r}_e,{\bf R}_p) =U({\bf r}_e) + U({\bf R}_p)\,\,
\end{equation}
describing the interaction of an electron ($q_e=-e$) and a proton ($q_p=e$) with field (1), respectively.
Passing to the coordinates of the center of mass ${\bf R}$ and relative motion ${\bf r}$ in the hydrogen atom
\begin{equation}
{\bf r}_e =(1-\frac{m_e}{M}){\bf r} + {\bf R} \approx {\bf r} + {\bf R}\,\,\,\,\,\,\, {\bf p}_e={\bf p}+\frac{m_e}{M}{\bf P}\approx {\bf p}\,\,
\end{equation}
\begin{equation}
{\bf R}_p = {\bf R} -\frac{m_e}{M}{\bf r} \approx {\bf R}\,\,\,\,\,\,\,\,\,\,\,\,\,\,\,\,\,\,\,\,\,\,\,\,\,\,\,\,\,\, {\bf p}_p=\frac{m_p}{M}{\bf P}-{\bf p}\approx {\bf P} - {\bf p}\,\,,
\end{equation}
with neglecting terms of the order of $M^{-1}=(m_e+m_p)^{-1}$, we finally obtain the interaction potential (4) of the hydrogen atom with the laser field (1) in the non-dipole approximation with an accuracy of the order of $c^{-2}$ and $M^{-1}$ (in the atomic units $\hbar=e^2=m_e=1$)
\begin{equation}
V({\bi r},{\bi R}) = V_1({\bi r}) + V_2({\bi r},{\bi R})\,\,,
\end{equation}
where
\begin{equation}
V_1({\bi r}) = E_0f(t)\{\cos(\omega t)x +\frac{1}{c}[\cos(\omega t)\hat{l}_y + \omega\sin(\omega t)xz]\}\,\,,
\end{equation}
and
\begin{equation}
V_2({\bi r},{\bi R}) = \frac{1}{c} E_0f(t)\{\cos(\omega t)[Z\hat{p}_x-X \hat{p}_z] + \omega\sin(\omega t)[xZ + zX]\}\,\,.
\end{equation}



\end{document}